\documentclass[useAMS,usenatbib,onecolumn]{mn2e}
\usepackage{graphicx}
\usepackage{pslatex}
\usepackage{natbib}
\usepackage{txfonts}

\newcommand{\vv}[1]{\bmath{#1}}
\newcommand{\mtr}[1]{\mathbfss{#1}}
\begin{document}

\title{YORP torque as the function of shape harmonics}

\author[Breiter and Michalska]
{
S\l awomir Breiter\thanks{Astronomical Observatory of A. Mickiewicz University,
 S\l oneczna 36, PL 60-286 Pozna\'n, Poland, breiter@amu.edu.pl},
Hanna Michalska\thanks{Astronomical Observatory of A. Mickiewicz University,
 S\l oneczna 36, PL 60-286 Pozna\'n, Poland, han@lab.astro.amu.edu.pl}
 }

 \pagerange{\pageref{firstpage}--\pageref{lastpage}}

\maketitle
\label{firstpage}

   \begin{abstract}
The second order analytical approximation of the mean YORP torque
components is given as an explicit function of the shape
spherical harmonics coefficients for a sufficiently regular minor body.
The results are based upon a new expression for the insolation function,
significantly simpler than in previous works. Linearized plane parallel
model of the temperature distribution derived from the insolation function
allows to take into account a nonzero conductivity.
Final expressions for the three average components of the YORP torque
related with rotation period, obliquity, and precession are
given in a form of Legendre series of the cosine of obliquity.
The series have good numerical properties and can be easily truncated according to the degree of
Legendre polynomials or associated functions, with first two
terms playing the principal role. The present version fixes the errors discovered in the
text that appeared in Monthly Notices RAS (388, pp. 297-944).
   \end{abstract}
   \begin{keywords}
   {radiation mechanisms: thermal---methods: analytical---celestial
   mechanics---minor planets, asteroids}
   \end{keywords}

\section{Introduction}

The Solar radiation affects rotation of minor bodies in the Solar system
by two principal mechanisms: direct pressure and thermal re-radiation.
The former has been shown to be of secondary importance \citep{NV:08a}, because
its average effect is negligible. But the significance of the thermal
re-radiation, known since the paper of \citet{Rub:00} under the name of
the YORP (Yarkovsky-O'Keefe-Radzievskii-Paddack) effect, is generally acclaimed and empirically
demonstrated \citep{Low:07,Tay:07,Kaasa:07}. A good account of the related problems can be found
in \citep{BVRN:06}.

The YORP torques acting on an irregular object are rather difficult to model.
Until the last year, only the results of numerical simulations were available,
with the torque values generated by summing the contributions from thousands of
facets of a triangulated surface \citep{VC:02,CV:04,Sch:07}. Although
\citet{Sch:07} provided a formula for the partially averaged YORP torque due to a single
small triangular facet in terms of the elliptic integrals, his formulation remains
of a semi-analytical type.
The only known general relations between the body shape and the YORP torque consisted
in the statement about the absence of YORP for spherical objects, and
the `windmill asymmetry' as the necessary condition for its appearance
\citep{Rub:00}. As it was shown by \citet{BMV:07}, if the thermal conductivity cannot be neglected,
the notion of the `windmill asymmetry' is wide enough to cover even the case of ellipsoids
of revolution.

The first step towards the explanation of the YORP torque
values in terms of an arbitrary body shape function was made in the recent paper by
\citet{NV:07}. The body shape was expressed in terms of spherical harmonics
and the general expressions for the mean component of the YORP torque
responsible for the rotation period variations were derived within the second
order approximation, i.e involving the products of shape harmonics coefficients.
The resulting expressions are cumbersome, involving the products
of Wigner functions for the dependence on obliquity, and tabulated coefficients
of power series in cosine of obliquity does not offer an explanation
for the notion of the YORP order introduced in the paper.
Recently, this line of research has been extended to the YORP torque component
responsible for variations in obliquity, with a new element -- the nonzero conductivity
influence \citep{NV:08}, but at the moment of submission we know only the preliminary
version of this article, so our comments about its content have to be rather vague.

Our present contribution, developed in parallel to the work of Nesvorn\'y and Vokrouhlick\'y,
is based upon similar patterns: shape description in terms of spherical harmonics,
a linear, 'plane parallel' temperature distribution model, and the second order
approximation. Yet, we believe that a less complicated formulation of the insolation
function, that we succeeded to obtain, leads to more convenient final expressions.
The Wigner functions play the role of a ladder that becomes unnecessary after
one climbs up to the final form of the mean YORP torque components, expressible in terms of
the Legendre polynomials and associated functions of order 1. Formulating the mean torque as Legendre series,
we introduce the notion of the YORP degree, complementary to the YORP order
of \citet{NV:07}, directly related to the degree of Legendre functions used, and helpful
when it comes to truncating the final expressions.

\section{General formula for the YORP torque}

Consider an infinitesimal, outwards oriented surface element $\mathrm{d}\vv{S}$ of a small
body, having temperature $T$. If the Lambert emission model
is assumed, the thermal radiation of this element gives rise to the
force \citep{BVRN:06}
\begin{equation}\label{df}
 \mathrm{d}\vv{f} = - \frac{2}{3}\frac{\varepsilon_\mathrm{t}\, \sigma}{v_\mathrm{c}}
 \,T^4\,\mathrm{d}\vv{S},
\end{equation}
where $\varepsilon_\mathrm{t}$ is the surface element emissivity, $\sigma$ is the Stefan-Boltzmann constant,
and $v_\mathrm{c}$ stands for the velocity of light. Acting on a surface element with the radius vector
from the centre of mass $\vv{r}$, the force produces an infinitesimal
torque $\mathrm{d}\vv{M} = \vv{r} \times \mathrm{d}\vv{f}$. Assuming the common emissivity for
all surface elements and integrating over the entire surface of the body, we obtain the net YORP torque
\begin{equation}\label{M}
    \vv{M} = - \frac{2}{3}\frac{\varepsilon_\mathrm{t}\, \sigma}{v_\mathrm{c}}
 \,\oint_S  T^4\,(\vv{r} \times \mathrm{d}\vv{S}).
\end{equation}
The temperature $T$ will be considered a continuous (but not necessarily smooth) function of the longitude and latitude of a surface element, and
of the Sun position.

Obviously, if the body is spherical, the vectors $\vv{r}$ and $\mathrm{d}\vv{S}$ are parallel,
so the torque $\vv{M}$ vanishes identically regardless of the temperature distribution function.
On the other hand, using elementary identities of the vector calculus, we can rewrite Eq.~(\ref{M}),
replacing the surface integral that involves
$\mathrm{d}\vv{S}$ by an integral over the volume of the considered body
\begin{equation}\label{M:1}
    \vv{M} =   \frac{2}{3}\frac{\varepsilon_\mathrm{t}\, \sigma}{v_\mathrm{c}}
 \,\int_V  \left[ \nabla \times ( T^4\,\vv{r} )\right]\, \mathrm{d}V = \frac{2}{3}\frac{\varepsilon_\mathrm{t}\, \sigma}{v_\mathrm{c}}
 \,\int_V  (\nabla T^4) \times \,\vv{r} \, \mathrm{d}V.
\end{equation}
This form is not more useful for the purpose of the present paper than the usual
Eq.~(\ref{M}), but it leads to a second basic fact about the YORP torque: $\vv{M}$
vanishes if the temperature distribution is isotropic, regardless of the body shape.
In other words, the mean value of the temperature over the surface is irrelevant for
the computation of YORP.

\section{Body shape model and insolation function}

\subsection{Surface equation in the body frame}

Let us introduce the reference frame $Oxyz$ with the origin at the centre of mass and
the axes aligned with the principal axes of inertia. In this `body frame'
we can express the object shape in the form of a parametric equation
\begin{equation}\label{shre}
    r = a + a \sum_{l \geq 1} \sum_{m=0}^{l} \Theta_l^m(\cos{\theta})
    \left[C_{l,m} \cos{m \lambda} + S_{l,m} \sin{m \lambda} \right],
\end{equation}
being the function of the colatitude $\theta$, longitude $\lambda$, and of the reference radius $a$.
In further discussion we will always use
\begin{equation}\label{uw}
    u = \cos{\theta}, \quad w = \sin{\theta} = \sqrt{1-u^2}.
\end{equation}
The normalized Legendre functions $\Theta_l^m(u)$ are defined in Appendix~\ref{sferfun}.
Normalized, dimensionless shape coefficients $C_{l,m}$ and $S_{l,m}$ will be assumed small
quantities of the first order. Their small values should allow the treatment by means of the
perturbation approach, and they should guarantee the convexity of figure, because no shadowing
is allowed in our model, as long as the Sun remains above the local tangent plane.

The real form of Eq.~(\ref{shre}) is instructive, but less advantageous in further
transformations. For this reason we
pass to the complex form of the real-valued equation for $r$
\begin{equation}\label{eq1}
    r = a + a \sum_{l \geq 1} \sum_{m=-l}^l f_{l,m} Y_{l,m}(u,\lambda),
\end{equation}
expressed in terms of spherical functions $Y_{l,m}$ (see Appendix~\ref{sferfun}),
and of complex shape coefficients
\begin{eqnarray}
    f_{l,0} & = &   C_{l,0}, \nonumber \\
    f_{l,m} & = &  (C_{l,m} - i\,S_{l,m})/2,
    \mbox{~~for $m > 0$},  \label{flm} \\
    f_{l,-m} & = & \left(-1\right)^m \,(C_{l,m} + i\,S_{l,m})/2 = (-1)^m \,f_{l,m}^\ast. \nonumber
\end{eqnarray}
Inverting Eqs.~(\ref{flm}) we find
\begin{equation}
% \nonumber to remove numbering (before each equation)
  C_{l,0} =  f_{l,0}, \quad
  C_{l,m} = \left( f_{l,m}+(-1)^m f_{l,-m} \right)  =   f_{l,m} + f^\ast_{l,m}   , \quad
  S_{l,m} = i\,\left( f_{l,m}-(-1)^m f_{l,-m} \right)  = i\,\left( f_{l,m} - f^\ast_{l,m} \right),
\end{equation}
in agreement with the properties (\ref{almcon}) of harmonic series.

For the sake of brevity, we introduce a symbol $\Psi$, so that
\begin{eqnarray}
    r & = & a\,(1+\Psi), \label{eq1:r}\\
    \Psi & = &  \sum_{l \geq 1} \sum_{m=-l}^l f_{l,m} Y_{l,m}(u,\lambda). \label{eq1:Psi}
\end{eqnarray}
$\Psi$ and its derivatives will be considered small quantities of the first order.

\subsection{Normal vector approximation}

In this paper we will frequently use a modified set of spherical coordinates with associated unit vectors
\begin{equation}\label{baza}
    \vv{\hat{e}}_r = \left(%
\begin{array}{c}
  w\, \cos{\lambda} \\ w\, \sin{\lambda} \\ u \\
\end{array}%
\right),
\quad
    \vv{\hat{e}}_\lambda = \frac{1}{w}\,\frac{\partial \vv{\hat{e}}_r}{\partial \lambda} =  \left(%
\begin{array}{c}
  -\sin{\lambda}  \\ \cos{\lambda} \\0 \\
\end{array}%
\right),
\quad
    \vv{\hat{e}}_u = w \,\frac{\partial \vv{\hat{e}}_r}{\partial u} = \left(%
\begin{array}{c}
  -u\, \cos{\lambda} \\ -u \, \sin{\lambda} \\ w \\
\end{array}%
\right),
\end{equation}
forming a right-handed orthonormal basis. Restricting the differentiation to the surface of the unit sphere $\mathcal{S}$, we
define a spherical surface gradient $\nabla_\mathrm{s}$ as
\begin{equation}\label{nabs}
    \nabla_\mathrm{s} f \equiv \frac{1}{w}\, \frac{\partial f}{\partial \lambda}\,\vv{\hat{e}}_\lambda
    + w\,\frac{\partial f}{\partial u}\,\vv{\hat{e}}_u.
\end{equation}

Given the shape function (\ref{eq1:r}), and setting $\vv{r}= r\,\vv{\hat{e}}_r$,  we can compute the outward normal vector on the body surface
\begin{equation}\label{Ndef}
    \vv{N} = \frac{\partial \vv{r}}{\partial \lambda} \times \frac{\partial \vv{r}}{\partial u}
    = r \left( r\, \frac{\partial \vv{\hat{e}}_r}{\partial \lambda} \times \frac{\partial \vv{\hat{e}}_r}{\partial u}
    + \frac{\partial r}{\partial u} \,\frac{\partial \vv{r}}{\partial \lambda} \times \vv{\hat{e}}_r +
    \frac{\partial r}{\partial \lambda} \, \vv{\hat{e}}_r \times \frac{\partial \vv{r}}{\partial u} \right)
    = a^2 \,(1+\Psi) \left( (1+\Psi)\, \vv{\hat{e}}_r - \frac{1}{w} \,\frac{\partial \Psi}{\partial
    \lambda}\, \vv{\hat{e}}_\lambda - w \,\frac{\partial \Psi}{\partial u}\,\vv{\hat{e}}_u\right) =
    a^2 \,(1+\Psi) \left( (1+\Psi)\, \vv{\hat{e}}_r - \nabla_\mathrm{s} \Psi\right).
\end{equation}
The normal vector $\vv{N}$ given above is relatively easy to handle, but the unit normal vector $\vv{\hat{n}}$  derived from
(\ref{Ndef}) has so complicated form, that we will use its linear approximation with respect to the shape coefficients
\begin{equation} \label{n:2}
\vv{\hat{n}} = \frac{\vv{N}}{N} \approx \vv{\hat{e}}_r- \frac{1}{w}\,\frac{\partial \Psi}{\partial \lambda}
    \,\vv{\hat{e}}_\lambda - w\,\frac{\partial \Psi}{\partial u}\, \vv{\hat{e}}_u
    = \vv{\hat{e}}_r - \nabla_\mathrm{s} \Psi.
\end{equation}
The first order approximation is sufficient for our purpose, because it will
be multiplied by a first order quantity inside the integral (\ref{M}).

\subsection{Solar position}

Although the shape of an object has been described in the body frame, we will temporarily use another
reference frame, where the description of the Solar position and of the terminator equation is
more convenient. Thus we introduce the $Ox' y' z'$ orbital frame, with the $Ox'$
axis pointing towards the Sun,  $Oz'$ axis directed along the orbital momentum
vector of the minor body, and the $Oy'$ axis completing the orthogonal, right-handed triad.

\begin{figure}
  % Requires \usepackage{graphicx}
  \begin{center}
  \includegraphics[width=8cm]{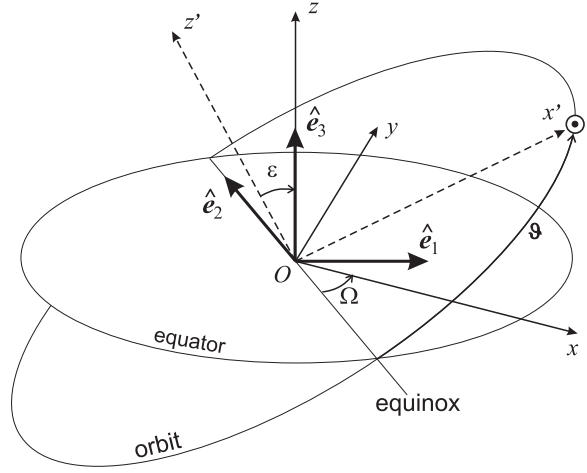}
  \end{center}
  \caption{Reference frames and orientation angles used in the
  paper (\textsf{corrected}).}\label{fig:1}
\end{figure}

In terms of the usual 3-1-3 Euler angles and rotation matrices $\mtr{R}_i $, the transformation from the body frame
$Oxyz$ to the orbital frame $Ox' y' z'$ requires (Fig.~\ref{fig:1})
\begin{equation}\label{rot1}
    \vv{r}' = \mtr{R}_3(\vartheta) \mtr{R}_1(\varepsilon)\mtr{R}_3(-\Omega)\, \vv{r}.
\end{equation}
However, a 3-2-3 rotation sequence is preferred in physics and we will use it in the present discussion.
As a matter of fact, the two sets of Euler angles are very closely related and
\begin{equation}\label{rot2}
    \vv{r}' = \mtr{R}_3(\psi) \mtr{R}_2(\varepsilon)
    \mtr{R}_3(\phi)\, \vv{r},
\end{equation}
is fully equivalent to Eq.~(\ref{rot1}) if we set\footnote{\textsf{corrected}}
\begin{equation}\label{equiv}
    \phi = -\Omega-\frac{\pi}{2},  \quad  \psi  = \vartheta + \frac{\pi}{2}.
\end{equation}
Assuming the circular, Keplerian motion approximation, we use
a constant obliquity angle $\varepsilon$, whereas $\Omega$ reflects the `diurnal' rotation,
and the argument of latitude $\vartheta$ is an angle that varies according to the `yearly' orbital motion.
The unit vector towards the Sun $\vv{\hat{n}}'_\odot$ and the vector normal to orbital plane $\vv{\hat{n}}'_\mathrm{o}$
are given by trivial expression in the orbital frame
\begin{equation}\label{ns1}
    \vv{\hat{n}}'_\odot = (1,0,0)^\mathrm{T} = \vv{\hat{e}}'_x, \quad \vv{\hat{n}}'_\mathrm{o} = (0,0,1)^\mathrm{T}=\vv{\hat{e}}_z'.
\end{equation}
Applying the inverse (transposed) rotation matrix (\ref{rot2}) we transform $\vv{\hat{n}}'_\odot = \vv{\hat{e}}_x'$
to the body frame, obtaining
\begin{equation}\label{sun:pos}
 \vv{\hat{n}}_\odot = (c\,\cos{\phi}\,\cos{\psi} - \sin{\phi}\sin{\psi})\,\vv{\hat{e}}_x
 + (c\,\sin{\phi}\,\cos{\psi} + \cos{\phi}\sin{\psi})\,\vv{\hat{e}}_y - s\,\cos{\psi}\,\vv{\hat{e}}_z.
\end{equation}
Throughout the text we use the symbols
\begin{equation}\label{csdef}
    c = \cos{\varepsilon}, \quad s = \sin{\varepsilon},
\end{equation}
related with the obliquity angle $\varepsilon$. Both $c$ and $s$ are considered constant and the motion of the Sun in the body frame
consists of the yearly motion described by $\psi$ and the apparent daily motion reflected in $\phi$.

\subsection{Insolation function}

\label{insol:sec}

The so-called insolation function is primarily defined in terms of the unit normal vector of the body surface $\vv{\hat{n}}$
and the unit vector directed to the Sun $\vv{\hat{n}}_\odot$ as
\begin{equation} \label{E:first}
\mathcal{E} = (1-A)\,\Phi\,\max{(0,\vv{\hat{n}}\cdot\vv{\hat{n}}_\odot)},
\end{equation}
where $A$ designates the Bond albedo and $\Phi = \Phi_0 a_\mathrm{o}^{-2}$ is the ratio of
the solar constant $\Phi_0 \approx 1366~\mathrm{W\, m^{-2}}$ to the square of the
orbital semi-major axis $a_\mathrm{o}$ expressed in astronomical units. $\mathcal{E}$ determines
the energy flux received by a given surface element of an illuminated body.

According to Eq.~(\ref{n:2}), we study an approximate insolation function involving
\begin{equation} \label{E:second}
 \vv{\hat{n}}\cdot\vv{\hat{n}}_\odot \approx \vv{\hat{e}}_r\cdot\vv{\hat{n}}_\odot - \frac{1}{w^2}\,\frac{\partial \Psi}{\partial \lambda}
    \,\frac{\partial ( \vv{\hat{e}}_r\cdot\vv{\hat{n}}_\odot )}{\partial \lambda}
    - w^2\,\frac{\partial \Psi}{\partial u} \frac{\partial ( \vv{\hat{e}}_r\cdot\vv{\hat{n}}_\odot )}{\partial u},
\end{equation}
because $ \vv{\hat{n}}_\odot $ is independent on $u$ and $\lambda$.

The function $\max{(0,\vv{\hat{n}}\cdot\vv{\hat{n}}_\odot)}$ is actually equivalent to a product
$H(\vv{\hat{n}}\cdot\vv{\hat{n}}_\odot) \,\vv{\hat{n}}\cdot\vv{\hat{n}}_\odot$, where $H(x)$ is the Heaviside
theta function. The values of  $H(\vv{\hat{n}}\cdot\vv{\hat{n}}_\odot)$ are: 1 on the sunny side of the body, 0 on the dark side,
and $\frac{1}{2}$ on the terminator. As it was demonstrated by \citet{NV:07}, replacing the actual terminator by the
terminator of a spherical body leads to an error of the second order. Thus, within the framework of the first order approximation,
we can replace $H(\vv{\hat{n}}\cdot\vv{\hat{n}}_\odot)$ by $H(\vv{\hat{e}}_r\cdot\vv{\hat{n}}_\odot)$. Moreover, thanks to the properties
of the Heaviside function, we can replace $H(y)(\mathrm{d}y/\mathrm{d}x)$ by $\mathrm{d}(H(y)\,y)/\mathrm{d}x$, so
\begin{eqnarray}
 \max{(0,\vv{\hat{n}}\cdot\vv{\hat{n}}_\odot)} & \approx & H(\vv{\hat{e}}_r\cdot\vv{\hat{n}}_\odot)\,\vv{\hat{e}}_r\cdot\vv{\hat{n}}_\odot
 - \frac{H(\vv{\hat{e}}_r\cdot\vv{\hat{n}}_\odot)}{w^2}\,\frac{\partial \Psi}{\partial \lambda}
    \,\frac{\partial ( \vv{\hat{e}}_r\cdot\vv{\hat{n}}_\odot )}{\partial \lambda}
    - H(\vv{\hat{e}}_r\cdot\vv{\hat{n}}_\odot)\,w^2\,\frac{\partial \Psi}{\partial u}
    \, \frac{\partial ( \vv{\hat{e}}_r\cdot\vv{\hat{n}}_\odot )}{\partial u} =
    \Xi  - \frac{1}{w^2}\,\frac{\partial \Psi}{\partial \lambda}
    \,\frac{\partial \Xi }{\partial \lambda}
    -  w^2\,\frac{\partial \Psi}{\partial u}
    \, \frac{\partial\Xi}{\partial u} = \nonumber \\
    & = &  \Xi - \left(\nabla_\mathrm{s} \Xi\right) \cdot \left(\nabla_\mathrm{s} \Psi\right),
\label{E:third}
\end{eqnarray}
where
\begin{equation}\label{Ksidef}
  \Xi =  H(\vv{\hat{e}}_r\cdot\vv{\hat{n}}_\odot)\,\vv{\hat{e}}_r\cdot\vv{\hat{n}}_\odot.
\end{equation}

In order to expand $\Xi$ in spherical harmonic series, we benefit from the rotational invariance of the scalar product
and consider $\Xi'$ in the orbital frame, where
\begin{equation}\label{Xiorb}
    \Xi' = H(\vv{\hat{e}}_r'\cdot \vv{\hat{e}}_x') \, \vv{\hat{e}}_r'\cdot \vv{\hat{e}}_x' =  H(w'\,\cos{\lambda'})\,w'\,\cos{\lambda'}.
\end{equation}
Using the rule (\ref{acoef}) we obtain $\Xi' = \sum_{l,k} \xi'_{l,k} Y_{l,k}(u',\lambda')$ with
the coefficients
\begin{eqnarray}
 \xi'_{l,k} & = & \int_{-1}^1 \mathrm{d}u' \int_{-\pi}^\pi H(w'\,\cos{\lambda'})\,w'\,\cos{\lambda'}\,Y^\ast_{l,k}(u',\lambda')\,\mathrm{d}\lambda' =
 \int_{-1}^1 \mathrm{d}u' \int_{-\frac{\pi}{2}}^{\frac{\pi}{2}} w'\,\cos{\lambda'}\,Y^\ast_{l,k}(u',\lambda')\,\mathrm{d}\lambda' = \nonumber \\
 & = &
  H_k \int_{-1}^1 w' \Theta_l^k(u') \,\mathrm{d}u' =  - \frac{H_k}{\sigma_{1,1}} \int_{-1}^1  \Theta_1^1(u') \Theta_l^k(u') \,\mathrm{d}u'
  = - \frac{H_k\,w_{1,l}^{1,k}}{\sigma_{1,1}}, \label{Ksi:1}
\end{eqnarray}
expressed in terms of the overlap integrals $w_{1,l}^{1,k}$ described in Appendix~\ref{OVG}, and coefficients $H_k$  defined
as
\begin{equation} \label{Hk}
 H_k = \int_{-\frac{\pi}{2}}^{\frac{\pi}{2}} \cos{L}\,e^{ikL} \, \mathrm{d} L,
\end{equation}
or, explicitly
\begin{equation}
H_k = H_{-k} = \left\{
\begin{array}{ccc}
    \left(1+(-1)^k\right) \frac{(-1)^{\frac{k}{2}+1}}{k^2-1} & \mbox{for} & k^2 \neq 1,\\
   \frac{1}{2}\,\pi & \mbox{for} & k^2=1, \\
\end{array}
  \right.
\end{equation}
with the particular case $H_0=2$, and $H_{2k+1} = 0 $ for all $|k| > 0$.

Note that, apart from $l=1$, only even degree $l$ and order $k$ terms will have nonzero coefficients $\xi'_{l,k}$, but we will not use this
information explicitly for the time being.

The transformation of $\Xi'$ back to the body frame is performed by means of the usual Wigner functions apparatus (see Appendix~\ref{Wigner}).
So we obtain
\begin{equation}\label{Xibf}
    \Xi = \sum_{l \geqslant 0} \sum_{k=-l}^l \sum_{m=-l}^l \xi'_{l,k} \,D^l_{m,k}(\phi,\varepsilon,\psi)\,Y_{l,m}(u,\lambda)
    =   \sum_{l \geqslant 0} \sum_{m=-l}^l \left[ - \sum_{k=-l}^l \frac{H_k\,w_{1,l}^{1,k}}{\sigma_{1,1}} \,D^l_{m,k}\right]\,Y_{l,m}
    = \sum_{l \geqslant 0} \sum_{m=-l}^l \xi_{l,m}\, Y_{l,m}.
\end{equation}
Throughout the main body of the article, we will use the convention, that $D^l_{m,k}$ without an explicit argument
always designates $D^l_{m,k}(\phi,\varepsilon,\psi)$, and $Y_{l,m}$ without an argument is a function of the body frame
coordinates $Y_{l,m}(u,\lambda)$.

If we substitute $\Xi$ from Eq.~(\ref{Xibf}) into (\ref{E:third}), we can write the insolation function as a sum
\begin{eqnarray}
  \mathcal{E} & \approx & \mathcal{E}^{(0)} + \mathcal{E}^{(1)}, \\
  \mathcal{E}^{(0)} &=& -(1-A)\,\Phi\,\sum_{l \geqslant 0} \sum_{k=-l}^l \sum_{m=-l}^l
   \frac{H_k\,w_{1,l}^{1,k}}{\sigma_{1,1}} \,D^l_{m,k}\,Y_{l,m},   \\
  \mathcal{E}^{(1)} & = & - (1-A)\,\Phi\, \sum_{l \geqslant 0} \sum_{m=-l}^l
  \sum_{j \geqslant 1} \sum_{k=-j}^j f_{j,k}\,\xi_{l,m}\,\left( \nabla_\mathrm{s} Y_{j,k} \right)
  \cdot \left( \nabla_\mathrm{s} Y_{l,m} \right),
\label{E1:first}
\end{eqnarray}
where we introduced the explicit form (\ref{eq1:Psi}) of the nonspherical shape function $\Psi$.

Using a vector calculus identity
for the scalar product of two gradients on the unit sphere
\begin{equation}\label{vec:id}
   \Delta_\mathrm{s}(F_1 F_2) = \nabla_\mathrm{s} \cdot \nabla_\mathrm{s}(F_1 F_2)=
   F_1\,\Delta_\mathrm{s} F_2+  2 (\nabla_\mathrm{s} F_1) \cdot (\nabla_\mathrm{s} F_2)
    + F_2\,\Delta_\mathrm{s} F_1,
\end{equation}
we achieve the transformation
\begin{equation}
  \left( \nabla_\mathrm{s} Y_{j,k} \right)
  \cdot \left( \nabla_\mathrm{s} Y_{l,m} \right)   =
    - \frac{1}{2}\,\left[Y_{j,k}\,\Delta_\mathrm{s} Y_{l,m} + Y_{l,m}\,\Delta_\mathrm{s} Y_{j,k} -\Delta_\mathrm{s}(Y_{j,k} Y_{l,m})\right]
     =  \frac{1}{2}\,\left[ \left(l\,(l+1) + j \,(j+1) \right)Y_{j,k}\, Y_{l,m}  + \Delta_\mathrm{s}(Y_{j,k} Y_{l,m})\right],
\label{trans:1}
\end{equation}
where we used the fact, that a spherical harmonic $Y_{l',m'}$ is the eigenfunction of the Laplacian operator on the unit sphere
$\Delta_\mathrm{s}$, with the eigenvalue $-l'(l'+1)$. Now, if we convert the products $Y_{j,k} Y_{l,m}$ into Clebsch-Gordan series
(\ref{Y:prod}), we obtain
\begin{equation}\label{trans:2}
   \left( \nabla_\mathrm{s} Y_{j,k} \right)
  \cdot \left( \nabla_\mathrm{s} Y_{l,m} \right)  =
      \frac{1}{2} \sum_{p,q} (-1)^{q} \left(l\,(l+1) + j \,(j+1)- p\,(p+1) \right)\,\mathcal{G}_{l,j,p}^{m,k,-q}\,Y_{pq},
\end{equation}
where the Gaunt coefficients $\mathcal{G}_{l,j,p}^{m,k,-q}$ are nonzero only if $q=m+k$, according to their properties
(\ref{G:prop}).

Thus the first order approximation of the insolation function is given by
\begin{equation}
\label{E:fin}
 \mathcal{E} \approx \sum_{l \geqslant 0} \sum_{m=-l}^l \left( \mathcal{E}^{(0)}_{l,m} + \mathcal{E}^{(1)}_{l,m} \right)\,Y_{l,m},
\end{equation}
where
\begin{eqnarray}
 \label{E0:fin}
  \mathcal{E}^{(0)}_{l,m} &=&  -(1-A)\,\Phi\, \sum_{k=-l}^l
   \frac{H_k\,w_{1,l}^{1,k}}{\sigma_{1,1}} \,D^l_{m,k} =  -(1-A)\,\Phi\, \sum_{k=-l}^l
   \frac{H_k\,w_{1,l}^{1,k}}{\sigma_{1,1}} \,d^l_{m,k}\,e^{-i(m\phi + k\,\psi)} , \\
 \label{E1:fin}
  \mathcal{E}^{(1)}_{l,m} &=& (-1)^{m}\, (1-A)\,\Phi\, \sum_{j \geqslant 1} \sum_{k=-j}^j  f_{j,k}\, \sum_{p \geqslant 0 }^{l+j}
   g_{p,j}^l\,\mathcal{G}_{p,j,l}^{m-k,k,-m} \sum_{q=-p}^p\frac{H_q\,w_{1,p}^{1,q}}{\sigma_{1,1}}\,d^p_{m-k,q}
   \,e^{-i\left((m-k)\phi + q\,\psi\right)}, \\
  g_{p,j}^l &= & \frac{1}{2}\,\left(p\,(p+1)+j\,(j+1)-l\,(l+1) \right).
\end{eqnarray}
Unless otherwise stated, a Wigner d-function $d^l_{m,k}$ without an explicit argument is $d^l_{m,k}(\varepsilon)$ defined
in Eq.~(\ref{smald}). Both Eqs.~(\ref{E0:fin}) and (\ref{E1:fin}) contain many terms that actually vanish in virtue
of the properties possessed by functions $H_k$, $w_{1,l}^{1,k}$, and $\mathcal{G}_{p,j,l}^{m-k,k,-m}$, but we
delay the task of purging the series until the last stage of the YORP torque derivation, preferring rather a compact
notation in intermediate transformations.

\section{Temperature model}

 If the body is treated as a perfectly nonconducting matter, with
the thermal conductivity $K =0,$ the temperature of a flat surface element is described by
\begin{equation}\label{Trub}
    T^4 = \frac{\mathcal{E}}{\varepsilon_\mathrm{t}\,\sigma},
\end{equation}
where $\mathcal{E}$ is the insolation function. This idealized temperature function leads to the models of YORP called Rubincam's
approximation \citep{Rub:00,VC:02}.

A common practice in modeling the YORP torques for bodies with nonzero coductivity,
is to assume a so-called `plane-parallel' temperature model \citep{CV:04, SchM:08, NV:08, Mysen:08}.
It is based on a one-dimensional (radial) heat diffusion equation
\begin{equation}\label{hde}
    \frac{\partial T}{\partial t} = \chi \, \frac{\partial^2 T}{\partial \zeta^2},
\end{equation}
where $\zeta$ measures the depth ($\zeta = 0$ on the surface, $\zeta > 0$ inside the body).
The coefficient $\chi$ is the ratio of the thermal conductivity $K$ to the product of the
density $\rho$ and the specific heat capacity $c_\mathrm{p}$
\begin{equation}\label{chi}
   \chi = \frac{K}{\rho \, c_\mathrm{p}}.
\end{equation}
The boundary conditions on the body surface are related with the conservation of energy, which implies
\begin{equation}\label{bc:1}
    \varepsilon_\mathrm{t} \, \sigma \, T^4 - K\,\left[ \frac{\partial T}{\partial \zeta}\right]_{\zeta=0}
    - \mathcal{E} = 0.
\end{equation}
The first term is the re-radiated energy, proportional to the fourth power of $T$, and
the product of the Boltzmann constant $\sigma$ and the surface emissivity $\varepsilon_\mathrm{t}$.
The second term is responsible for the heat transport towards the  body's centre,
and the last term is the insolation function $\mathcal{E}$, providing the incoming radiation energy.
Moreover, we search only the steady state solution that is a quasi-periodic function of time, and such
that $\lim_{\zeta \rightarrow \infty} \frac{\partial T}{\partial \zeta} = 0$.

Nonlinear boundary conditions are linearized around the constant value $T_0$, given by
\begin{equation}
\label{bcl:0}
  \varepsilon_\mathrm{t}\,\sigma\,T_{0}^4  = \mathcal{E}^{(0)}_{00} = \frac{(1-A)\,\Phi}{4},
\end{equation}
leading to the linear model
\begin{equation}\label{Tlin}
    T \approx T_0 + \tilde{T},
\end{equation}
where $\tilde{T}$ is given by
\begin{eqnarray}
  4\,\varepsilon_\mathrm{t}\,\sigma\,T_{0}^3\,\tilde{T} -
  K\,\left[ \frac{\partial \tilde{T}}{\partial \zeta} \right]_{\zeta=0} = \mathcal{E} - \mathcal{E}^{(0)}_{00}.
\label{bcl:2}
\end{eqnarray}

A more detailed description of this model can be found in \citep{BFV:03} or \citep{Vok:98}.
For our purpose, we simply state an operational rule, that each term of the insolation function $\mathcal{E}$
having a form
\[
   B_{pq}(u,\lambda)\, e^{i \,(p \phi + q\,\psi)},
\]
creates a term $\tilde{T}_{pq}$ defined by
\begin{equation}\label{gensol}
   4\,\varepsilon_\mathrm{t}\,\sigma\,T_{0}^3\, \tilde{T}_{pq} = B_{pq}(u,\lambda)\, K_{pq}\, e^{i \,(p \phi + q \psi + \varphi_{pq})},
\end{equation}
where, assuming a constant rotation rate $\omega = - \dot{\phi}$,
and a circular orbit with  mean motion $n_\mathrm{s} = \dot{\psi}$, the thermal conductivity
effects are described in terms of
\begin{eqnarray}
 K_{pq} & = &  \frac{1}{\sqrt{1 + 2 \, F_{pq}+ 2\,F_{pq}^2 }}, \\
 \sin{\varphi_{pq}} & = &   \mathrm{sgn}(p \omega - q\,n_\mathrm{s})\, F_{pq}\,K_{pq}, \\
 \cos{\varphi_{pq}} & = &  \left(1+ F_{pq} \right)\,K_{pq}, \\
\label{Fmk}
    F_{p,q} & = &
  \frac{\rho\,c_\mathrm{p}}{4\,\varepsilon_\mathrm{t}\,\sigma\,T_0^3}
  \,\sqrt{\frac{|p \omega - q n_\mathrm{s}|\,\chi}{2}}.
\end{eqnarray}
Note, that $K_{-p,-q} = K_{p,q}$, but $\varphi_{-p,-q} = -\varphi_{pq}$. If the second subscript is 0, we will
use an abbreviated form
\begin{equation}\label{abbr}
    K_{p0} = K_p, \quad \varphi_{p0} = \varphi_p,
\end{equation}
with important special cases $K_0=1$, and $\varphi_0 = 0$.

If we apply the rule (\ref{gensol}) to the insolation function
series (\ref{E:fin}-\ref{E1:fin}) and split $\tilde{T}$ into
the spherical part $T^{(0)}$ and non-spherical correction $T^{(1)}$, we find
\begin{equation}\label{tem}
 \varepsilon_\mathrm{t} \sigma  T^4  \approx
 \varepsilon_\mathrm{t} \sigma  T_0^4 + 4  \varepsilon_\mathrm{t} \sigma  T_0^3 \,
 \left(T^{(0)} + T^{(1)} \right)  = \frac{(1-A)\,\Phi}{4} +
   \sum_{l \geqslant 1} \sum_{m=-l}^l
 t^{(0)}_{l,m}  \,Y_{l,m}(u,\lambda) +
 \sum_{l \geqslant 0} \sum_{m=-l}^l
   t^{(1)}_{l,m} \,Y_{l,m}(u,\lambda),
\end{equation}
with
\begin{eqnarray}
\label{t0lm}
  t^{(0)}_{l,m}
 & = & - (1-A)\,\Phi\,\sum_{k=-l}^{l}
    \frac{w_{1,l}^{1,k}\,H_{k}}{\sigma_{1,1}}
    \,K_{m,k}\,  d^{l}_{m,k} \, e^{-i\,(m \phi + k \psi + \varphi_{m,k})}, \\
\label{t1lm}
  t^{(1)}_{l,m} &=&   (1-A)\,\Phi  \sum_{j \geqslant 1} \sum_{p = 0 }^{l+j} \sum_{q=-p}^p
    \sum_{k=-j}^j f_{j,k}\,
  (-1)^{m}\, g_{pj}^l\,\mathcal{G}_{p,j,l}^{m-k,k,-m}\,
  \frac{w_{1,p}^{1,q}\,H_{q}}{\sigma_{1,1}}
  \, d^{p}_{m-k,q}  \,K_{m-k,q} \, e^{-i\,((m -k) \phi + q \psi + \varphi_{m-k,q})}.
\end{eqnarray}
This temperature model is accurate up to the first power of the shape coefficients
$f_{j,k}$ not only because of the approximate insolation function used to derive it,
but also because of $T_0$ defined according to the spherical model.

\section{Torque integration over the surface}

\subsection{Approximate integrand}

The integral (\ref{M}) involves the factor $(\vv{r} \times \mathrm{d}\vv{S})$. In terms
of the modified spherical coordinates in the body frame,
\begin{equation}\label{dS}
    \mathrm{d}\vv{S} = \vv{N}\,\mathrm{d}u\,\mathrm{d}\lambda,
\end{equation}
where the normal vector $\vv{N}$ is given by Eq.~(\ref{Ndef}).
Substituting Eq.~(\ref{eq1:r}), we find
\begin{equation}
  \vv{r} \times \vv{N} = a^3\,(1+\Psi)^2\,\vv{\hat{e}}_r \times  \left[ (1+\Psi)\, \vv{\hat{e}}_r - \nabla_\mathrm{s} \Psi\right]
  = - a^3\,(1+\Psi)^2\,\vv{\hat{e}}_r \times  \nabla_\mathrm{s} \Psi
 = - i\, a^3\,(1+\Psi)^2 \vv{L}^\mathrm{s} \Psi,
\end{equation}
where $\vv{L}^\mathrm{s} = - i \vv{\hat{e}}_r \times  \nabla_\mathrm{s}$ is the angular momentum
operator \citep{Jack,BieLo} reduced to the unit sphere. Up to the second order, we
can approximate $\vv{r} \times \vv{N}$ as
\begin{equation}\label{rtN}
    \vv{r} \times \vv{N} \approx - i\, a^3\,(1+2\,\Psi) \vv{L}^\mathrm{s} \Psi = a^3 (1+2 \Psi)\, \vv{F}.
\end{equation}
According to the action of $\vv{L}^\mathrm{s}$ on spherical harmonics
in the  complex, body frame-fixed basis
\begin{equation}\label{baza:pm}
   \vv{\hat{e}}_- = \frac{1}{\sqrt{2}}\,\left(\vv{\hat{e}}_x-i \vv{\hat{e}}_y \right),
   \qquad \vv{\hat{e}}_+ =  \vv{\hat{e}}_-^\ast
   =   \frac{1}{\sqrt{2}}\,\left(\vv{\hat{e}}_x+i \vv{\hat{e}}_y \right),
   \qquad \vv{\hat{e}}_0 = \vv{\hat{e}}_z,
\end{equation}
the vector $\vv{F}= - i \vv{L}^\mathrm{s} \Psi$, is a sum
\begin{equation}
\vv{F} =  \sum_{l \geqslant 1} \sum_{m=-l}^l f_{l,m}\, \vv{F}_{l,m},
\end{equation}
where the vector coefficients
\begin{equation} \label{Fdef:1}
\vv{F}_{l,m} = -i\,\left(%
  \sigma_{l,m}^-  \, Y_{l,m+1} \, \vv{\hat{e}}_-
 +  \sigma_{l,m}^+ \,Y_{l,m-1}\, \vv{\hat{e}}_+
 + m\,Y_{l m}\,\vv{\hat{e}}_0 \right) ,
\end{equation}
involve auxiliary symbols
\begin{equation}\label{sig:pm}
\sigma_{l,m}^- = \sqrt{\frac{(l-m)(l+m+1)}{2}}= \sigma_{l,-m}^+,
\qquad \sigma_{l,m}^+ = \sqrt{\frac{(l+m)(l-m+1)}{2}}   = \sigma_{l,-m}^-.
\end{equation}
Note that from the point of view of a usual scalar product $\vv{\hat{e}}_\pm \cdot \vv{\hat{e}}_0=0$,
and $\vv{\hat{e}}_0 \cdot \vv{\hat{e}}_0=1$, but $\vv{\hat{e}}_+ \cdot \vv{\hat{e}}_+=
\vv{\hat{e}}_-  \cdot \vv{\hat{e}}_- =0$, and $\vv{\hat{e}}_- \cdot \vv{\hat{e}}_+ =1$.

According to the conclusions drawn from Eq.~(\ref{M:1}), we can drop
$l=m=0$ terms from our temperature model (\ref{tem}), because they have no effect
on the YORP torque. Hence, we will use
\begin{equation}\label{tem:red}
 \varepsilon_\mathrm{t} \sigma  T^4 = \sum_{l \geqslant 1} \sum_{m=-l}^l
 (t^{(0)}_{l,m} + t^{(1)}_{l,m})\,Y_{l,m},
\end{equation}
in further considerations. Thus we have reached the stage, where the problem of the YORP torque is reduced to
an integral over the surface of the unit sphere
\begin{equation}\label{M:7}
    \vv{M} \approx - \frac{2\,a^3}{3\,v_\mathrm{c}}
 \,\int_{-1}^1 \mathrm{d}u \int_{0}^{2\pi}  \sum_{l \geqslant 1} \sum_{m=-l}^l
 \left[ t^{(0)}_{l,m} \, \vv{F} +2 \, t^{(0)}_{l,m} \,\Psi\,\vv{F} + t^{(1)}_{l,m}\,\vv{F}\right]\,Y_{l,m} \,\mathrm{d}\lambda,
\end{equation}
where the first term in the square bracket is a first order YORP torque and
remaining two define the second order torque. Any second order term $t^{(2)}_{l,m}$
in temperature function,
depending on the products of shape coefficients, would result in a third order
contribution after being multiplied by the first order vector $\vv{F}$. This
justifies the use of the first order approximation of $T^4$ in the second order
YORP torque solution.

\subsection{Integrated torque in the body frame}

The first order part of Eq.~(\ref{M:7}) is
\begin{equation}\label{M4:1}
    \vv{M}^{(1)} = - \frac{2\,a^3}{3\,v_\mathrm{c}} \sum_{l \geqslant 1} \sum_{m = -l}^l
    \sum_{j \geqslant 1} \sum_{k = -j}^j f_{l,m}\, t^{(0)}_{j,k}\,
    \,\int_{-1}^1 \mathrm{d} u \, \int_{0}^{2\,\pi} Y_{j,k}\,\vv{F}_{l,m}\, \mathrm{d}\lambda.
\end{equation}
If we substitute $Y_{j,k} = (-1)^k Y_{j,-k}^\ast$ and use the orthogonality of
spherical functions (\ref{orto}), we easily find
\begin{equation}\label{M4:3}
    \vv{M}^{(1)} =  - \frac{2\,i\,a^3}{3\,v_\mathrm{c}}
    \sum_{l \geqslant 1} \sum_{m = -l}^l f^\ast_{l,m}\,
    \left[\sigma_{l,m}^-\, t^{(0)}_{l,m+1}\,  \vv{\hat{e}}_+
    + \sigma_{l,m}^+\, t^{(0)}_{l,m-1}\,\vv{\hat{e}}_-
    + m\,t^{(0)}_{l,m}\,\vv{\hat{e}}_0 \right].
\end{equation}

In the second order we have $\vv{M}^{(2)} = \vv{M}^{(02)} + \vv{M}^{(11)}$, with
\begin{equation}\label{M21}
    \vv{M}^{(11)} = - \frac{2\,a^3}{3\,v_\mathrm{c}} \sum_{l \geqslant 1} \sum_{m = -l}^l
    \sum_{j \geqslant 1} \sum_{k = -j}^j f_{l,m}\, t^{(1)}_{j,k}\,
    \,\int_{-1}^1 \mathrm{d} u \, \int_{0}^{2\,\pi} Y_{j,k}\,\vv{F}_{l,m}\, \mathrm{d}\lambda,
\end{equation}
resembling $\vv{M}^{(1)}$, and
\begin{equation}\label{M22:1}
    \vv{M}^{(02)} = - \frac{4\,a^3}{3\,v_\mathrm{c}} \sum_{l \geqslant 1} \sum_{m = -l}^l
    \sum_{j \geqslant 1} \sum_{k = -j}^j \sum_{p \geqslant 1} \sum_{q = -p}^p
    f_{l,m}\,f_{p,q} t^{(0)}_{j,k}\,
    \,\int_{-1}^1 \mathrm{d} u \, \int_{0}^{2\,\pi} Y_{j,k}\,Y_{p,q}\,\vv{F}_{l,m}\, \mathrm{d}\lambda.
\end{equation}

Of course, the final form of $\vv{M}^{(11)}$ is similar to Eq.~(\ref{M4:3}), with all
$t^{(0)}_{j,k}$ replaced by $t^{(1)}_{j,k}$,
\begin{equation}\label{M21:2}
    \vv{M}^{(11)} =  - \frac{2\,i\,a^3}{3\,v_\mathrm{c}}
    \sum_{l \geqslant 1} \sum_{m = -l}^l f^\ast_{l,m}\,
    \left[\sigma_{l,m}^-\, t^{(1)}_{l,m+1}\,  \vv{\hat{e}}_+
    + \sigma_{l,m}^+\, t^{(1)}_{l,m-1}\,\vv{\hat{e}}_-
    + m\,t^{(1)}_{l,m}\,\vv{\hat{e}}_0 \right],
\end{equation}
whereas $\vv{M}^{(02)}$ looks different, but still the integral of three
spherical functions is directly given in terms of the Gaunt integrals (\ref{G}).
In our case, we obtain
$\mathcal{G}_{l,p,j}^{m,q,k}$, $\mathcal{G}_{l,p,j}^{m,q,k+1}$,
and $\mathcal{G}_{l,p,j}^{m,q,k-1}$, that are nonzero only for $m+q+k$ equal
to 0, $-1$ and $1$ respectively (see Appendix~\ref{OVG}). This
property helps to remove one sum, and
\begin{equation}
    \vv{M}^{(02)}   =   \frac{4\,i\,a^3}{3\,v_\mathrm{c}} \sum_{l,j,p \geqslant 1}  \sum_{m = -l}^l
    \sum_{q = -p}^p
    f_{l,m}\,f_{p,q} \,
\left(%
\, \sigma_{l,m}^-\,\mathcal{G}_{l,p,j}^{m+1,q,-m-q-1}\,t^{(0)}_{j,-m-q-1}\,\vv{\hat{e}}_-
 + \sigma_{l,m}^+ \,\mathcal{G}_{l,p,j}^{m-1,q,-m-q+1}\,t^{(0)}_{j,-m-q+1}\,\vv{\hat{e}} _+
  + m\,\mathcal{G}_{l,p,j}^{m,q,-m-q}\,t^{(0)}_{j,-m-q} \vv{\hat{e}}_0
\right).
\label{M22:2}
\end{equation}
Further restrictions of summation range, implied by properties (\ref{G:prop}), will be introduced later.

The equations derived in this section provide a tool to compute the complete second-order approximation
of the YORP torque
\begin{equation}\label{Mtot}
    \vv{M} = \vv{M}^{(1)}+ \mu\,\vv{M}^{(02)} + \vv{M}^{(11)}
\end{equation}
if only the spherical harmonics expansion of the temperature distribution function is known
up to the first order in shape coefficients. We introduce a symbol $\mu=1$ as a marker
that will help us to locate the contribution of the $\vv{M}^{(02)}$, neglected by
\citet{NV:08}, in the final expressions.

\section{Elements of dynamics}

Given an object on a circular orbit, rotating around its principal axis of the maximum inertia, we can describe its
dynamics under the action of a torque $\vv{M}$ in terms of four differential equations related
to the 3-1-3 Euler angles\footnote{\textsf{corrected}} $\Omega = -\phi - \pi/2$, $\varepsilon$, $\vartheta = \psi - \pi/2$,
and to the angular rotation rate $\omega$:
\begin{eqnarray}
  \dot{\omega} &=&  \frac{\vv{M} \cdot \hat{\vv{e}}_3 }{J_3}, \label{eqm:1} \\
  \dot{\varepsilon} & = & \frac{\vv{M}\cdot \hat{\vv{e}}_1}{
  \omega\,J_3}, \label{eqm:2} \\
  \dot{\vartheta} & = & \dot{\psi} ~=~ n_\mathrm{s} - \frac{\vv{M}\cdot \hat{\vv{e}}_2}{
  \omega\,J_3\,s}, \label{eqm:3} \\
  \dot{\Omega} & = & - \dot{\phi} ~=~
 \omega - \frac{c\,\vv{M}\cdot \hat{\vv{e}}_2}{
  \omega\,J_3\,s} ~=~
  \omega + c\,(\dot{\vartheta} - n_\mathrm{s}), \label{eqm:4}
\end{eqnarray}
where $J_3$ is the maximum moment of inertia, $n_\mathrm{s}$ is the orbital mean motion and, as usually, $c =\cos{\varepsilon}$,
$s = \sin{\varepsilon}$.
The three unit vectors in the above equations define an inertial reference frame attached
to the body's equator-equinox system:\footnote{\textsf{corrected}}
$\hat{\vv{e}}_1$ -- opposite to the projection of
$\vv{\hat{e}}_z'$ (normal to the orbital plane) on the equatorial plane $Oxy$,
$\hat{\vv{e}}_2$ -- opposite to the vernal equinox (i.e. directed along $\vv{\hat{e}}_z' \times \vv{\hat{e}}_z$),
and $\hat{\vv{e}}_3$ -- directed to the north pole.
The components of this right-handed, orthonormal basis in the body frame are given by
\begin{equation}\label{vecs}
\begin{array}{l}
\hat{\vv{e}}_1  =
   -\cos{\phi}\,\vv{\hat{e}}_x
   -\sin{\phi}\,\vv{\hat{e}}_y = - \frac{1}{\sqrt{2}}\,\left[ e^{-i\phi}\, \vv{\hat{e}}_+
   +  e^{i\phi}\,\vv{\hat{e}}_- \right],\\
\hat{\vv{e}}_2   =
   \sin{\phi}\,\vv{\hat{e}}_x - \cos{\phi}\,\vv{\hat{e}}_y
   = \frac{i}{\sqrt{2}} \,\left[ e^{-i\phi} \, \vv{\hat{e}}_+  -
    e^{i\phi} \, \vv{\hat{e}}_- \right], \\
    \hat{\vv{e}}_3  =  \vv{\hat{e}}_z = \vv{\hat{e}}_0. \\
\end{array}
\end{equation}

\section{Mean values of the YORP torque projections}

The YORP torques belong to the class of very weak, dissipative factors that may produce significant effects only
through their mean values that act systematically. If we exclude any commensurability between $n_\mathrm{s}$ and $\omega$, we can identify
the averaging with respect to time with
\begin{equation}\label{aver}
    \langle M_j \rangle = \frac{1}{4 \pi^2} \int_{0}^{2\pi} \int_{0}^{2\pi} M_j \,\mathrm{d}\phi\,\mathrm{d}\psi,
\end{equation}
rejecting all purely periodic terms of $\phi$ and $\psi$ in
\begin{equation}\label{Mj}
    M_j = \vv{M} \cdot \vv{\hat{e}}_j, \quad j = 1,2,3.
\end{equation}

\subsection{Spin component $M_3$}

According to Eqs.~(\ref{eqm:1}), the mean component
$\langle M_3 \rangle =\vv{M}\cdot \vv{\hat{e}}_3$ plays the
principal role in the YORP effect, being responsible for the systematic
change in rotation period.  Besides, it also can be described with a simpler set
of formulae than $M_1$ or $M_2$. For these reasons we consider $M_3$ prior to
the remaining components.

\subsubsection{First order}

The first order terms in $M_3$, to be labeled $M_{3}^{(1)}$, result
from $\vv{M}^{(1)}$ defined in Eq.~(\ref{M4:3}). But
\begin{equation}\label{M4:31}
 \left\langle M_3^{(1)} \right\rangle = \left\langle \vv{M}^{(1)} \cdot \vv{\hat{e}}_3  \right\rangle =
  - \frac{2\,i\,a^3}{3\,v_\mathrm{c}}
    \sum_{l \geqslant 1} \sum_{m = -l}^l  f^\ast_{l,m}\, m\,\left\langle t^{(0)}_{l,m} \right\rangle = 0,
\end{equation}
because according to Eq.~(\ref{t0lm}), $\langle t^{(0)}_{l,m} \rangle =0$ for all $m \neq 0$,
whereas for $m=0$, we have $ m\,\langle t^{(0)}_{l,m} \rangle =0$. Thus, as demonstrated by
Nesvorn\'y and Vokrouhlick\'y (\citeyear{NV:07,NV:08}), the spin-up or spin-down of an asteroid is a second order effect, involving the products
of the shape coefficients.

\subsubsection{Second order}

The second order part of $M_3$ will consist of  terms derived from $\vv{M}^{(11)}$ and $\vv{M}^{(02)}$
defined in Eqs.~(\ref{M21:2}) and (\ref{M22:2}), where we substitute $t^{(0)}_{l,m}$ from
Eq.~(\ref{t0lm}) and $t^{(1)}_{l,m}$ from Eq.~(\ref{t1lm}). We begin with $M_3^{(11)}$
and after the substitution of temperature coefficients (\ref{t1lm}) with $k=m$, and $q=0$,
we readily obtain the mean value
\begin{equation}
  \left\langle M_3^{(11)} \right\rangle = \left\langle \vv{M}^{(11)} \cdot \vv{\hat{e}}_3 \right\rangle =
   - \frac{2\,i\,a^3}{3\,v_\mathrm{c}}
    \sum_{l \geqslant 1} \sum_{m = -l}^l  f^\ast_{l,m}\, m\,\left\langle t^{(1)}_{l,m} \right\rangle =
    - i\,\alpha  \, \sum_{l,p,j,m}
  (-1)^m\,m\,f^\ast_{l,m}\,f_{j,m}
    \frac{w_{1,p}^{1,0}\,H_{0}}{\sigma_{1,1}} \,K_{0,0}\,g_{p,j}^l\,\mathcal{G}_{l,j,p}^{-m,m,0}\,  d^{p}_{0,0} \, e^{-\,i \,\varphi_{0,0}},
\end{equation}
where
\begin{equation}\label{alf:def}
  \alpha  =   \frac{2\,a^3\,(1-A)\,\Phi}{3\,v_\mathrm{c}}.
\end{equation}
Recalling that $\varphi_{0,0}=0$, $K_{0,0}=1$, $H_0=2$, and
$w_{1,p}^{1,0}=0$ when $p$ is odd, we reduce  $\left\langle M_3^{(11)} \right\rangle$ to
\begin{equation}\label{wyn:01}
  \left\langle M_3^{(11)} \right\rangle =
  - i\,2\,\alpha  \sum_{l,j,p \geqslant 1} \sum_{m=-l}^l
  (-1)^m\,m\,f^\ast_{l,m}\,f_{j,m}
     \,g^l_{2p,j}\,\mathcal{G}_{l,j,2p}^{-m,m,0}\,  W_p\,  d^{2p}_{0,0},
\end{equation}
using the special case of the overlap integral $W_p$ defined in Eq.~(\ref{W2p}).

In agreement with the results of \citet{NV:07}, we find that the remaining  term $  M_3^{(02)}$
has no systematic effect on the spin rate. Indeed,
\begin{equation} \label{wyn:020}
  \left\langle M_3^{(02)} \right\rangle = i\,\frac{4\,a^3}{3\,v_\mathrm{c}} \sum_{l,j,p,m,q} m f_{l,m} f_{p,q}
  \, \mathcal{G}_{l,p,j}^{m,q,-m-q} \left\langle t^{(0)}_{j,-m-q}\right\rangle =
  i\, 4\, \alpha
  \sum_{l,j,p \geqslant 1} \sum_{m = -l}^l m \, (-1)^m\,  f^\ast_{l,m}\,f_{j,m}\, \mathcal{G}_{l,j,2p}^{-m,m,0}\,  W_p \,d^{2p}_{0,0}=0,
\end{equation}
thanks to the symmetries between the terms with the indices $(l,j,m)=(q_1,q_2, \pm q_3)$ and $(q_2,q_1, \pm q_3)$.
For similar reasons we can actually replace $g_{2p,j}^l$ in Eq.~(\ref{wyn:01}) by
$g_{0,j}^l$.

Thus, the complete mean YORP torque $\langle M_3 \rangle$ is given by
\begin{equation}\label{trans}
    \left\langle M_3 \right\rangle = \left\langle M_3^{(11)}  \right\rangle =
    - i\,2\,\alpha\,\sum_{l,j,p \geqslant 1} \sum_{m = -l}^l m \, (-1)^m\,  f^\ast_{l,m}\,f_{j,m}\,
   g^l_{0,j} \, \mathcal{G}_{l,j,2p}^{-m,m,0}\,  W_p \,d^{2p}_{0,0}.
\end{equation}

Equation~(\ref{wyn:01}) can be converted into an explicitly real form if we combine the positive and negative values of
the summation index $m$ and use the parity properties of shape coefficients, Wigner d-functions, and Gaunt coefficients.
The result, with vanishing terms dropped and with optimized summation with respect to $l$ and $j$, becomes
\begin{equation}
  \left\langle M_3 \right\rangle = - 2\,\alpha
   \sum_{l  \geqslant 1} \sum_{ j \geqslant 1} \sum_{m=1}^l
  (-1)^m\,m\,j\,\left(2l+2j+1\right)\,\left(C_{l,m} S_{l+2j,m} - S_{l,m} C_{l+2j,m}\right)
   \sum_{p = j}^{l+j}
    W_p \,\mathcal{G}_{l,l+2j,2p}^{-m,m,0}\,d^{2p}_{0,0}.
\label{wyn:02}
\end{equation}

Two remarkable properties known from previous works \citep{CV:04,NV:07, Sch:07,SchM:08} are worth noting:
$\left\langle M_3 \right\rangle$ is an even function of $(\pi/2 - \varepsilon)$, and it is
independent on thermal conductivity. The former property is a direct consequence of the fact,
that $d^{2p}_{00}$ is actually
the Legendre polynomial $P_{2p}(\cos{\varepsilon})$. A good argument for the independence of
$\left\langle M_3 \right\rangle$ on conductivity can be found in \citet{NV:08}.

\subsection{Obliquity component $M_1$}

\label{S:obl}

\subsubsection{First order}

Using Eq.~(\ref{M4:3}), we project $\vv{M}^{(1)}$ on $\vv{\hat{e}}_1$ defined
by Eq.~(\ref{vecs}) in terms of $\vv{\hat{e}}_\pm$. Then, recalling that
$\vv{\hat{e}}_\pm \cdot \vv{\hat{e}}_0 = \vv{\hat{e}}_+ \cdot \vv{\hat{e}}_+=
\vv{\hat{e}}_- \cdot \vv{\hat{e}}_-=0$, and $\vv{\hat{e}}_+ \cdot \vv{\hat{e}}_-=1$,
we find that the average value of the first order terms is given by
\begin{equation}\label{MM:1}
    \left\langle M_{1}^{(1)} \right\rangle =   i\,  \frac{\sqrt{2} \,a^3}{3\,v_\mathrm{c}}
    \sum_{l \geqslant 1} \sum_{m = -l}^l f^\ast_{l,m}\, \left(
     \sigma_{l,m}^+\, \left\langle   t^{(0)}_{l,m-1}\,e^{-i\phi}  \right\rangle
     + \sigma_{l,m}^-\, \left\langle   t^{(0)}_{l,m+1}\,e^{i\phi}  \right\rangle
     \right).
\end{equation}
Substitution of the spherical temperature function from Eq.~(\ref{t0lm}),
followed by the rejection of purely periodic terms (in both components it amounts to
setting $m=k=0$), leads to
\begin{eqnarray}\label{MM:2}
    \left\langle M_{1}^{(1)} \right\rangle & = &
     - i\,  \frac{ \alpha}{\sqrt{2} \,\sigma_{1,1}}
    \sum_{l \geqslant 1}   f^\ast_{l,0}\,
     w_{1,l}^{1,0}\,H_{0}\left(
  K_{1,0}\, \sigma_{l0}^-\,  d^{l}_{1,0} \, e^{-i\, \varphi_{1,0}}
 + K_{-1,0}\, \sigma_{l0}^+\,  d^{l}_{-1,0} \, e^{-i\, \varphi_{-1,0}}\right).
\end{eqnarray}
Using the particular values of the subscripted symbols, namely
\[
\varphi_{-1,0} = - \varphi_1, \quad K_{-1,0}=K_{1}, \quad H_0=2, \quad d^{l}_{-1,0} = - d^{l}_{1,0},
\quad \sigma_{l,0}^+ = \sqrt{\frac{l\,(l+1)}{2}},\quad f^\ast_{l,0}=C_{l,0},
\]
we find the first order contribution in an explicitly real form
\begin{equation}\label{MM:3}
    \left\langle M_{1}^{(1)} \right\rangle =-  \frac{2 \, \alpha}{\sigma_{1,1}}\,K_1 \,\sin{\varphi_1}
    \sum_{l \geqslant 1}   C_{l,0}\, \sqrt{l\,(l+1)}  \,
     w_{1,l}^{1,0} \,  d^{l}_{1,0},
\end{equation}
depending only on zonal harmonics coefficients $C_{l0}$. Moreover, the properties of overlap integrals
(\ref{over}) imply that only the even degree zonals remain in the final formula, so -- introducing the
special function $W_l$ from Eq.~(\ref{W2p}) -- we finally obtain
\begin{equation}\label{MEQ:F}
    \left\langle M_{1}^{(1)} \right\rangle = - 2 \, \alpha \,K_1 \sin{\varphi_1}
    \sum_{l \geqslant 1}   C_{2l,0}\, \sqrt{2l\,(2l+1)}  \,
     W_l\,  d^{2l}_{1,0}.
\end{equation}
This expression confirms the presence of the YORP effect in obliquity even for
such regular objects as spheroids, provided they have nonzero conductivity
\citep{BMV:07}.

\subsubsection{Second order}

In the second order we first consider $M_{1}^{(11)}$. According to Eq.~(\ref{M21:2}),
the departure point is similar to $M_{1}^{(1)}$, i.e.,
\begin{equation}
    \left\langle M_{1}^{(11)} \right\rangle =   i\,  \frac{\sqrt{2} \,a^3}{3\,v_\mathrm{c}}
    \sum_{l \geqslant 1} \sum_{m = -l}^l f^\ast_{l,m}\, \left(
     \sigma_{l,m}^+\, \left\langle   t^{(1)}_{l,m-1}\,e^{-i\phi}  \right\rangle
     + \sigma_{l,m}^-\, \left\langle   t^{(1)}_{l,m+1}\,e^{i\phi}  \right\rangle
     \right).
\end{equation}
although the temperature coefficient $t^{(1)}_{l,m-1}$ from Eq.~(\ref{t1lm}) is more complicated.
The $\phi$ and $\psi$ independent part, resulting from $k=m$ and $q=0$, is
\begin{equation} \label{MM:4}
 \left\langle M_1^{(11)} \right\rangle    =   i\,\sqrt{2}\,\alpha \,K_1
 \sum_{l,j,p,m}  (-1)^m\,f_{l,m}^\ast f_{j,m}\,g^l_{2p,j}  \, \left(
 \sigma^+_{l,m}\, \mathcal{G}_{j,l,2p}^{m,-m+1,-1}\,e^{i\varphi_1} -
 \sigma^-_{l,m}\, \mathcal{G}_{j,l,2p}^{m,-m-1,1}\,e^{-i\varphi_1}
 \right) \, W_p\,d^{2p}_{1,0} .
\end{equation}

The treatment of $M_1^{(02)}$ is comparable to $M_1^{11}$. The result
becomes quite similar to $M_1^{(11)}$, because in this case
 Eqs.~(\ref{M22:2}) and (\ref{t0lm}) lead to
\begin{eqnarray} \label{M22:temp}
   \left\langle M_{1}^{(02)} \right\rangle  & = &
 - i\,\frac{2 \sqrt{2}\,a^3}{3\,v_\mathrm{c}} \sum_{l,j,p,m,q}
 f_{l,m} f_{p,q} \left( \sigma^-_{l,m} \mathcal{G}_{l,p,j}^{m+1,q,-m-q-1}
 \left\langle e^{-i\phi}\,t^{(0)}_{j,-m-q-1}\right\rangle
 +\sigma^+_{l,m} \mathcal{G}_{l,p,j}^{m-1,q,-m-q+1}
 \left\langle e^{i\phi}\,t^{(0)}_{j,-m-q+1}\right\rangle \right) = \nonumber \\
  & = &     i\,2 \sqrt{2}\, \alpha\,K_1
    \sum_{l,j,p,m}   (-1)^m \,f_{l,m} f^\ast_{p,m}\,\left(
    \sigma^+_{l,m}\,\mathcal{G}_{l,p,2j}^{m-1,-m,1}\,e^{-i\varphi_1}
   - \sigma^-_{l,m}\,\mathcal{G}_{l,p,2j}^{m+1,-m,-1}\,e^{i\varphi_1}\right)\,W_{j}
    \,d^{2j}_{1,0}.
\end{eqnarray}
In order to add Eq.~(\ref{M22:temp}) to (\ref{MM:4}), we manipulate the former,
 interchanging the indices $(l,p,j)$ , permuting the indices in the Gaunt coefficients, and
 making use of $f_{l,-m}= (-1)^m\,f^\ast_{l,m}$, and $\sigma^-_{l,-m} = \sigma^+_{l,m}$. The resulting total second order torque is
\begin{equation} \label{MM:5}
 \left\langle M_{1}^{(2)}  \right\rangle =
 \left\langle M_{1}^{(11)}+ \mu M_{1}^{(02)}  \right\rangle    =
 i\,\sqrt{2}\,\alpha\,K_1
 \sum_{l,j,p,m}  (-1)^m\,f_{l,m}^\ast f_{j,m}\,
 \left(2\mu-g^l_{2p,j}\right) \,\left(
  \sigma^-_{l,m}\,  \mathcal{G}_{j,l,2p}^{m,-1-m,1}
   \,e^{-i\varphi_1}
 - \sigma^+_{l,m}\,  \mathcal{G}_{j,l,2p}^{m,1-m,-1}
   \,e^{i\varphi_1} \right)\,W_{p}\,d^{2p}_{1,0}.
\end{equation}
The marker $\mu=1$ introduced in Eq.~(\ref{Mtot}) helps us to trace the influence of
the $M_1^{(02)}$.

After a tedious procedure of combining the terms with $(l,j,m) = (q_1,q_2,\pm q_3)$
and $(l,j,m) = (q_2,q_1,\pm q_3)$, we obtain the explicitly real form
\begin{eqnarray}
  \left\langle M_1^{(2)} \right\rangle  & = & \alpha\,K_1 \,\cos{\varphi_1}
  \sum_{l \geqslant 1} \sum_{j \geqslant 1} \sum_{m = 1}^l (-1)^m \,\left( C_{l,m} S_{l+2j,m} - C_{l+2j,m}S_{l,m} \right)
  \, g^l_{0,l+2j}    \sum_{p =j}^{j+l} \sqrt{2}
  \left( \sigma^+_{l,m}\,\mathcal{G}_{l,l+2j,2p}^{m-1,-m,1}
  - \sigma^-_{l,m}\, \mathcal{G}_{l,l+2j,2p}^{m+1,-m,-1}   \right)\,W_p\,\,d^{2p}_{1,0} \nonumber \\
 & & +  \alpha\,K_1\,\sin{\varphi_1}
  \sum_{l \geqslant 1} \sum_{j \geqslant 0} \sum_{m = 0}^l (-1)^m \,\left( C_{l,m} C_{l+2j,m} + S_{l,m} S_{l+2j,m} \right)
  \, \frac{2-\delta_{0,j}}{2-\delta_{0,m}}
   \sum_{p = \max{(1,j)}}^{j+l}  \frac{2 g^l_{2p,l+2j} g^{l+2j}_{2p,l}- 4p (2p+1)\,\mu}{\sqrt{2p\,(2p+1)}}\,\mathcal{G}_{l,l+2j,2p}^{m,-m,0} \,W_p\,\,d^{2p}_{1,0}.
\label{MB:fin}
\end{eqnarray}
Using the recurrence relation (\ref{Grec}) we were able to simplify the sum of two
Gaunt integrals, but it did not help for their difference, which explains asymmetry
of the two parts of Eq.~(\ref{MB:fin}). Notice, that $\mu$ is present only in the terms
factored by $\sin{\varphi_1}$, i.e., in the part that does vanish in the Rubincam's
approximation.

\subsection{Precession component $M_2$}

Probably, the YORP induced precession will meet rather limited interest, being negligible when
compared to the precession resulting from much stronger gravitational torques.
Fortunately, the similarity between $\vv{\hat{e}}_1$ and $\vv{\hat{e}}_2$ resulted
in a very simple rule:  it suffices to apply a formal substitution
$(\sin{\varphi}_1,\,\cos{\varphi_1}) \rightarrow (-\cos{\varphi}_1,\,\sin{\varphi_1})$   to turn
$M_1^{(1)}$ or $M_1^{(2)}$ into
$M_2^{(1)}$  or $M_2^{(2)}$ respectively. Of course, this observation is
based on the complete derivation, that we omit for the sake of brevity,
but it can also be deduced from the equations of \citet{Sch:07} or \citet{SchM:08}.

\subsection{Summary}

\label{Sumar}

In order to collect the final equations for the average YORP torque in a possibly compact form,
revealing their actual structure, we first introduce few auxiliary symbols
\begin{eqnarray}
  S_{l,m}^j &=&   C_{l,m}S_{l+2j,m} - S_{l,m} C_{l+2j,m} ,  \\
  C_{l,m}^j &=&  C_{l,m}C_{l+2j,m} + S_{l,m} S_{l+2j,m},
\end{eqnarray}
and
\begin{equation}\label{sc12}
      s_1 =  K_1 \sin{\varphi_1}, \quad   c_1 =  K_1 \cos{\varphi_1},
\end{equation}
that should not be confused with $c=\cos{\varepsilon}$, and $s=\sin{\varepsilon}$.

Further, we note that only the special cases of Wigner d-functions are required: $d^{2p}_{0,0}$ and $d^{2p}_{1,0}$,
which are simply related with the Legendre polynomials $P_{2p}(c)$ and Legendre functions $P^1_{2p}(c)$, respectively.
We will also limit the summation according to a practical assumption, that shape harmonics are known only up to
some finite degree and order $N$.
Thus, using Eq.~(\ref{dl0k}), we conclude this section with the following set of formulae:
\begin{eqnarray}
\left(%
\begin{array}{c}
  \left\langle M_1 \right\rangle  \\
  \left\langle M_2 \right\rangle \\
\end{array}%
\right)
&=&
 - 2\,\alpha\,\left(%
\begin{array}{c}
  s_1  \\
  -c_1 \\
\end{array}%
\right) \sum_{l = 1}^{E[N/2]} C_{2l,0}\, W_l\, P^1_{2l}(c) \nonumber \\
 & &  +  \alpha\,\left(\begin{array}{c}
  s_1  \\
  -c_1 \\
\end{array}%
\right)
    \sum_{l = 1}^{N}
     \sum_{j = 0}^{E\left[\frac{N-l}{2}\right]} \sum_{m = 0}^l  C_{l,m}^j  \sum_{p=\max{(1,j)}}^{l+j} U^\mathrm{c}_{l,m,j,p}\,P^1_{2p}(c)
     +
 \alpha\,\left(\begin{array}{c}
  c_1  \\
  s_1 \\
\end{array}%
\right)
    \sum_{l = 1}^{N-2}
     \sum_{j = 1}^{E\left[\frac{N-l}{2}\right]} \sum_{m = 1}^l   S_{l,m}^j  \sum_{p=j}^{l+j} U^\mathrm{s}_{l,m,j,p}\,P^1_{2p}(c),     \\
  \left\langle M_3 \right\rangle &=&    \alpha
    \sum_{l = 1}^{N-2}
     \sum_{j = 1}^{E\left[\frac{N-l}{2}\right]} \sum_{m = 1}^l S_{l,m}^j  \sum_{p=j}^{l+j} V_{l,m,j,p} \,P_{2p}(c),
\end{eqnarray}
where\footnote{\textsf{corrected}}
\begin{eqnarray}
     V_{l,m,j,p}  & = & -2\,(-1)^m\,j\, m\,(2l+2j+1)\,\mathcal{G}_{l,l+2j,2p}^{m,-m,0}\,W_p, \\
 U^\mathrm{c}_{l,m,j,p}  & = & (-1)^m \frac{2-\delta_{0j}}{2-\delta_{0m}}\,
\left( \frac{ g^l_{2p,l+2j}\,g^{l+2j}_{2p,l}}{p\,(2p+1)} - 2 \mu  \right)
 \,\mathcal{G}_{l,l+2j,2p}^{m,-m,0}\,W_p, \\
 U^\mathrm{s}_{l,m,j,p} & = & \frac{(-1)^m\,j\,(2l+2j+1)}{ \sqrt{2p\,(2p+1)}}
 \mathcal{H}_{l,m,j,p}\,W_p, \\
\mathcal{H}_{l,m,j,p} & = & \sqrt{(l+m)(l-m+1)}\,\mathcal{G}_{l,l+2j,2p}^{m-1,-m,1}
 - \sqrt{(l-m)(l+m+1)}\,\mathcal{G}_{l,l+2j,2p}^{m+1,-m,-1},
\end{eqnarray}
$E[x]$ is the entier function, rounding down to the nearest integer, and $\delta_{j,k}$ is the Kronecker delta function.

The coefficients $W_p$ are quickly decreasing in absolute value from $W_0 = - \sqrt{\pi}/4$, with $\lim_{p \rightarrow \infty} W_p=0$.
They can be evaluated by means of the simple recurrence
\begin{equation}\label{Wp:recur}
    W_{p} = \sqrt{\frac{4p+1}{4p-3}}\,\frac{(2p-1)(2p-3)}{2p\,(2p+2)}\,W_{p-1}.
\end{equation}
The Gaunt coefficients present in the final expressions are defined in terms of the Wigner 3-j symbols in Eq.~(\ref{G3j}), so
\begin{equation}\label{G0:3j}
 \mathcal{G}_{l,l+2j,2p}^{m,-m,0} = \sqrt{\frac{ (2l+1)(2l+4j+1)(4p+1)}{4\pi}}
 \left[%
\begin{array}{ccc}
  l & l+2j & 2p \\
  0 & 0 & 0 \\
\end{array}%
\right] \,  \left[%
\begin{array}{ccc}
  l & l+2j & 2p \\
  m & -m & 0 \\
\end{array}%
\right],
\end{equation}
and
\begin{eqnarray}
 \mathcal{H}_{l,m,j,p} =  \sqrt{\frac{ (2l+1)(2l+4j+1)(4p+1)}{2\pi}}
 \left[%
\begin{array}{ccc}
  l & l+2j & 2p \\
  0 & 0 & 0 \\
\end{array}%
\right]  \left( \sigma^+_{l,m}\,\left[%
\begin{array}{ccc}
  l & l+2j & 2p \\
  m-1 & -m & 1 \\
\end{array}%
\right]   - \sigma^-_{l,m}\, \left[%
\begin{array}{ccc}
  l & l+2j & 2p \\
  m+1 & -m & -1 \\
\end{array}%
\right]  \right). & &
\end{eqnarray}
The 3-j symbols and Gaunt coefficients admit a number of recurrence relation and their
computation is a routine task in physics (eg. \citet{LuLu:98}, \citet{Xu:96}, \citet{Seb:98}). A reliable software for the 3-j coefficients
can be found in the open source SHTOOLS\footnote{available at \texttt{http://www.ipgp.jussieu.fr/\~{ }wieczor/SHTOOLS/SHTOOLS.html}}
package, although major computer algebra systems implement the 3-j symbols as one of standard functions.
The values of $\mathcal{G}_{l,l+2j,2p}^{m,-m,0}$ and $\mathcal{H}_{lmjp}$ decrease with the growing $p$,
but they do not reveal any monotonicity with respect to $l$, $n$, or $j$.

The presence of even degree Legendre polynomials of $c$ in $ \left\langle M_3 \right\rangle$
and of Legendre functions $P^1_{2p}(c)$ in  $\langle M_{1,2} \rangle (\varepsilon)$
explains the symmetries well known from \citep{Rub:00} or \citep{CV:04}:
$\langle M_3 \rangle (\varepsilon) = \langle M_3 \rangle (\pi-\varepsilon)$ and
$\langle M_1 \rangle (\varepsilon) = - \langle M_1 \rangle (\pi-\varepsilon)$.
Similarly, $\langle M_2 \rangle (\varepsilon) = - \langle M_2 \rangle (\pi-\varepsilon)$, although the division of $M_2$
by $s$ in the right hand sides of Eqs.~(\ref{eqm:3}) and (\ref{eqm:4}) reverts this symmetry and the effect on precession is
an even function of the angle $(\frac{\pi}{2} - \varepsilon)$.

In the Rubincam's approximation, when we neglect the conductivity, $s_1=0$ and $c_1=1$. This simplification has no effect
on the spin component $\left\langle M_3 \right\rangle$, in agreement with the results
of
\citet{CV:04,SchM:08,Mysen:08} and \citet{NV:08}.

\section{Legendre series form and truncation rules}

If the Legendre functions in the formulae of Section~\ref{Sumar} are expressed in terms of the powers of $c$,
we obtain exactly the same final expressions for spin component $\left\langle M_3 \right\rangle$ as \citet{NV:07}.
Inspecting the coefficients of such expansions in \citep{NV:07} one finds no rule that might
help to select the leading terms, responsible for the major part of the YORP torque. The situation is much
better in the present formulation, because the monotonicity properties of $W_p$ alone (even fortified in the
products $\mathcal{G}_{l,l+2j,2p}^{m,-m,0}\,W_p$ or $ \mathcal{H}_{l,j,m,p}\,W_p$) suggests to attach the major importance
to the terms with the lowest $p$ values. In this context, we can justify the partition of the YORP torque
terms into subsequent `orders' of $j$, proposed by \citet{NV:07}, because the index $j$ establishes the lower
bound on the range of $p$. Truncating with respect to $j$ is also consistent with the approach of
\citet{Mysen:08}, who imposed it at the beginning of his derivation, using an insolation function
that actually is restricted to the first two Legendre polynomials  $P_1$ and $P_2$, with the effect of
$P_1$ disappearing after the averaging.

The sums presented in the previous section can be easily rearranged from the form
attaching to each shape coefficients combination a sum of Legendre functions, to an
alternative form of Legendre series with a sum of coefficients attached to each Legendre function.
Simple manipulations lead to
\begin{eqnarray}
\label{M1:final}
\left\langle M_1 \right\rangle & = &  \alpha  \sum_{q = 1}^{N} \left( s_1\,X^1_q+s_1 X^2_q + c_1 Z_q \right)\,P^1_{2q}(c), \\
\label{M2:final}
\left\langle M_2 \right\rangle & = &  \alpha  \sum_{q = 1}^{N} \left( - c_1\,X^1_q - c_1 X^2_q + s_1 Z_q \right)\,P^1_{2q}(c),\\
\label{M3:final}
\left\langle M_3 \right\rangle & = &    \alpha \sum_{q = 1}^{N-1} A_q \,P_{2q}(c),
\end{eqnarray}
with the coefficients given as the sums
\begin{equation}
X^1_q   =    -2\,C_{2q}\,W_q, \quad
X^2_q   =   \sum_{j = 0}^{q} \sum_{l = l_1}^{N-2j} \sum_{m= 0}^{l} C_{l,m}^j \,U^\mathrm{c}_{l,m,j,q}, \quad
Z_q  =  \sum_{j = 1}^{q} \sum_{l = l_1}^{N-2j} \sum_{m=1}^{l} S_{l,m}^j \,U^\mathrm{s}_{l,m,j,q}, \quad
A_q   =  \sum_{j = 1}^{q} \sum_{l = l_1}^{N-2j} \sum_{m=1}^{l} S_{l,m}^j \, V_{l,m,j,q},
\end{equation}
where the lower summation bound is $l_1 = \max{(1,q-j)}$.

These Legendre series suggest to introduce a notion of the `YORP degree', complementary to the `YORP order'
proposed by \citet{NV:07}. A term of the YORP degree $q$ is a sum of terms with YORP orders $j \leqslant q$.
Truncating the YORP series one should first decide on the maximum degree $q_\mathrm{max}$, and then either maintain
the maximum order $j$ naturally bounded by the current value of $q$, or additionally impose some smaller value of $j_\mathrm{max}$.
According to our experience, a decent approximation level is reached already at $q_\mathrm{max}=2$,
so a qualitative pattern of the $\varepsilon$ dependence is established in the second degree
truncated series.\footnote{Coefficients files up to $N=50$ are available from the first author (SB) upon the request.}

\section{Test Applications}

\subsection{General remarks}

The basic assumptions of our solution are similar to the ones of \citet{NV:07,NV:08}:
almost spherical homogeneous bodies with constant albedo and
small thermal penetration depth combined with low conductivity. The shape restriction
is quite severe, ruling out many irregularly shaped objects like (433)~Eros. A good
account of this difficulty was given in Section 8 of \citep{NV:07}. Our experience
confirms the statement of \citet{NV:07} that qualitatively correct results can be obtained
even for such irregular bodies, although the main cause of discrepancies is not the magnitude
of the shape coefficients as such, but rather the occurrence of major concavities
in the figure of a considered body that generates large shadow zones treated as illuminated in
analytical insolation function models. So, we test our solution on a spheroid,
that is obviously convex, and on the asteroid 1998~$\mathrm{KY}_{26}$ -- a reasonably regular object.

\subsection{Spheroids}

As a first test we confronted the present results with YORP torque formulae for spheroids provided
by \citet{BMV:07}. Although the results for spheroids used a simplistic thermal lag model of `delayed Sun',
the final formulae from Sect.~3.5 of \citet{BMV:07} can be easily generalized to a more realistic
thermal model by replacing $\sin{\delta}$ with $s_1$ of the present paper. Then, with the shape model
\begin{equation}\label{sfero:1}
    C_{2,0} \approx - \frac{2}{3}\,\sqrt{\frac{\pi}{5}}\,\left(   e^2  + \frac{11}{21}\,e^4\right),
    \qquad
    C_{4,0} \approx \frac{2 \sqrt{\pi}}{35}\,e^4,
\end{equation}
derived for an oblate spheroid
\begin{equation}\label{sfero:2}
    r = \frac{a_\mathrm{e}\,\sqrt{1-e^2}}{\sqrt{1-e^2\,w^2}},
\end{equation}
with the semi-axis $a_\mathrm{e}$ related to the mean radius $a$  by the series in eccentricity
$e = \sqrt{1-c_\mathrm{e}^2\,a^{-2}_\mathrm{e}}$
\begin{equation}\label{sfero:3}
 a \approx a_\mathrm{e}\,\left(1 - \frac{e^2}{6} - \frac{11\,e^4}{120}\right),
\end{equation}
we find from Eq.~(\ref{M1:final})
\begin{equation}\label{sfero:4}
    \left\langle M_1 \right\rangle = \alpha\,s_1 \left( X_1^1 \,P_2^1(c) + X_2^1 \,P_4^1(c) + X_1^2\,P_2^1(c) \right) \approx
  - \frac{2 (1-A)\,\Phi\,a^3}{3\,v_\mathrm{c}}\,\frac{\pi\,s\,c}{128}\,s_1\,\left(16\,e^2 + (7+5c^2 )\,e^4 \right).
\end{equation}
On the other hand, according to \citet{BMV:07}
\begin{equation}\label{sfero:5}
 \left\langle M_1 \right\rangle \approx  - \frac{\pi}{12}\,\frac{ (1-A)\,\Phi\,a_\mathrm{e}^3}{ v_\mathrm{c}}\,s\,c\,\sin{\delta}
 \left(e^2 + \left(\frac{1}{4} - \frac{5}{16}\,s^2\right) \,e^4 \right).
\end{equation}
Replacing $\sin{\delta}$ by $s_1$ and substituting  Eq.~(\ref{sfero:3}) we obtain a perfect agreement between
Eqs.~(\ref{sfero:4}) and (\ref{sfero:5}).

The comparison is meaningful only up to $e^4$, because $C_{2,0}^2 = O(e^4)$ marks the limit of our present
second order approximation. For the same reasons we dropped $C_{6,0} = O(e^6)$ and higher degree zonal coefficients.

\subsection{Asteroid $1998~\mathrm{KY}_{26}$}

\begin{figure}
  \includegraphics{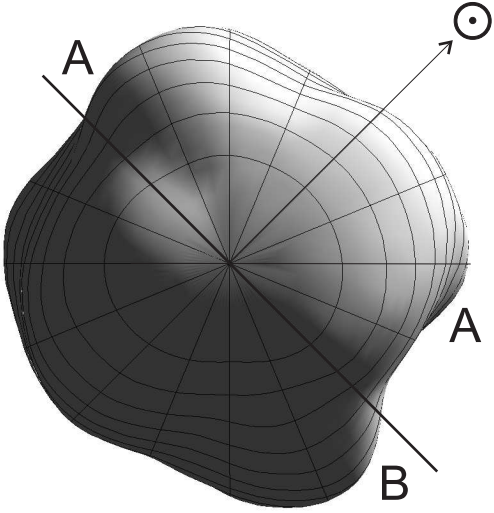}\\
  \caption{Asteroid 1998 KY26 seen from the north pole with a simulated lighting from the equator. The regions
  where our insolation function model fails are labeled A (self-shadowing) and B (terminator cuts off a
  sunlit region).}\label{fig:2}
\end{figure}

\begin{figure}
  \includegraphics[width=\linewidth]{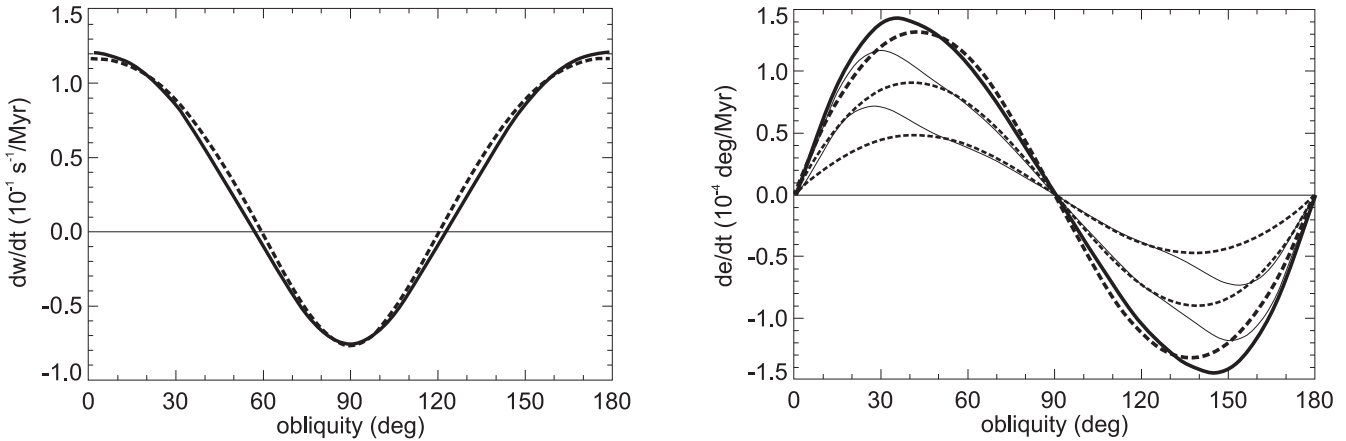}\\
  \caption{Present solution (continuous lines) for $1998~\mathrm{KY}_{26}$ and the numerical model of  \citet{CV:04} (dashed lines).
  Thick lines refer to the conductivity $K=0$. Thin lines present the results for $K=0.001$ and $K=0.01~\mathrm{W/(m\,K)}$. }\label{fig:3}
\end{figure}

From the collection of numerically simulated mean YORP torques presented by \citet{CV:04}, we selected
the asteroid 1998~$\mathrm{KY}_{26}$, considered also by \citet{Sch:07} and
\citet{SchM:08}. The same object was used by Nesvorn\'y and Vokrouhlick\'y \citeyear{NV:07,NV:08} as an example
for their second order model. The authors were quite optimistic about this body, and indeed, a side view presented in their paper looks like
well fitting the assumption of the theory. However, the polar view (Fig.~\ref{fig:2}) reveals
a source of possible problems when the obliquity $\varepsilon$ is close to $0$ or $\pi$.
Some regions, qualified as sunlit by the spherical terminator criterion are actually in shadow,
whereas some sunlit regions are neglected and considered dark. Apparently, these two effects cancel out in the numerical simulation,
but the analytical model tends to sharpen their influence.

Nevertheless, we constructed the spherical harmonics model up to degree and order $N=26$ using the least squares
adjustment on the grid of 2048 surface points determined by \citet{Ostro:99}, reduced to the reference frame of centre of mass
and principal axes of inertia. The standard deviation of our harmonics set on the grid points was $0.2~\mathrm{mm}$,
quite satisfactory for an object with the mean radius $a=13.1913~\mathrm{m}$.
Interested rather in comparison than in the dynamics of the asteroid itself, we used
the same, partially fictitious values of physical parameters that were applied by \citet{CV:04}: rotation period $6~\mathrm{hr}$,
density $\rho = 2.5~\mathrm{g/cm^3}$, orbital semi-axis $a_\mathrm{o} = 2.5~\mathrm{au}$, heat capacity $c_\mathrm{p}=680~\mathrm{J/(kg\,K)}$,
and albedo $A=0$. The maximum moment of inertia, that we derived during the reduction to the principal axes,
was $J_3 = 1.7273 \times 10^{9} \mathrm{kg\,m^2}$.

The agreement of $\dot{\omega}$ computed from Eq.~(\ref{M3:final}) shown in Fig.~\ref{fig:3} with the results of \citet{CV:04} is as good as the one
obtained by \citet{NV:07}. It is not surprising, because we expanded our expressions in power series of $c$ and compared them
with Table~I of \citet{NV:07}, concluding a perfect agreement of the numerical coefficients.  Yet, when we pass to
the obliquity variations $\dot{\varepsilon}$, we find much more distortion with respect to smooth curves taken from \citet{CV:04},
although the orders of magnitude and qualitative dependence on obliquity and thermal conductivity may be qualified as correct.
In our opinion, the degraded accuracy at small values of $\varepsilon$ or $(\pi-\varepsilon)$ should be attributed to the
north-south directed elevated patterns in the topography of $1998~\mathrm{KY}_{26}$. We have observed, that the departure
from the sine-like curve of \citet{CV:04} strongly depends on the maximum degree and order $N$ of the accounted shape harmonics.
As a matter fact, a sufficiently inaccurate shape model, with the elevated regions smoothed out, gives better shapes of the
$\dot{\varepsilon}$ curve, albeit producing a smaller amplitude.

\section{Conclusions}

Although we used the same assumptions and general strategy as Nesvorn\'y and Vokrouhlick\'y
(\citeyear{NV:07,NV:08}), we have obtained a significantly
simpler form of the final expressions of the mean YORP torque. In particular, the final formulae are expressed in terms of
Legendre polynomials and associated functions, as compared to the products of Wigner functions that can be found
in \citet{NV:07}. The advantage comes from the way we expand the insolation function -- more straightforward than
the one exposed in Appendix~B of \citet{NV:07}. Our formulation is well suited for the partition into the terms
of primary and secondary influence according to the YORP degree, directly related with the degree of Legendre functions
present in the final expressions. Moreover, it has significantly better numerical properties
than the power series of $c$.
 We confirm the basic facts established in recent years, like the independence
of the spin related torque component on conductivity \citep{CV:04,NV:08} within the plane parallel temperature model, or the
presence of YORP effect in obliquity for the bodies with rotational symmetry \citep{BMV:07}.
The explicit appearance of $P_2(c)$ in Eq.~(\ref{M3:final}) and the remarks
about the properties of $W_p$, neatly explain the fact that the spin related YORP
torques tend to vanish close to the zeros of $P_2(c)$ \citep{NV:07}.
Similarly we can predict, that if the obliquity component $\langle M_1\rangle$
has zeros different than $c=0$ or $s=0$ (implied by each Legendre function $P_{2p}^1(c)$),
the new zeros should be located in the domain $ 40^\circ < \varepsilon < 60^\circ $
(or $120^\circ < \varepsilon < 140^\circ$) implied by $P_2^1(c) \pm P_4^1(c) = 0$.
This conjecture is confirmed by a large sample of Gaussian
random spheres simulated in \citep{CV:04}, as well as by the observed
`Slivan states' in the Koronis family \citep{Slivan,Vok:03}.

The assumptions about the insolation model seem to be the main source of divergence between
numerical models and a completely analytical formulation. In this context, the approach taken by
\citet{Mysen:08} seems to be a good compromise: in his paper the insolation function
is derived from numerical simulation and then incorporated into an analytical treatment.
As usually, however, such semi-analytical approach breaks the direct link between
the shape coefficients and the YORP torque magnitude. It is a hard duty of an analytical
solution to show such link, even if the final relation becomes as cumbersome as the one
derived here.

Notably, the YORP effect in the Rubincam's approximation is a second order
effect in terms of the shape coefficients. Which of the harmonics products will play
the leading role, depends on the particular body shape, because there is no
monotonicity of the numerical coefficients accompanying a particular harmonics product.
It is easier to point out nonsignificant harmonics: for example, the spin torque
$\left\langle M_3 \right\rangle$ is independent on the zonal harmonics coefficients $C_{l,0}$.
Although the presence of the nonzero conductivity brings in the first-order terms to the
$\left\langle M_1 \right\rangle$ and $\left\langle M_2 \right\rangle$, in most cases they
will be less significant than the second order contribution, because $s_1$ -- factoring the zonal coefficients
-- is typically much smaller
than $c_1$. Thus the zonal harmonics coefficients are generally of minor importance, unless
the body has a highly regular shape, close to a rotationally invariant solid.

The modular structure of the present work allows a modification of particular elements of the
solution without restarting the entire work from the scratch. Our next goal is the application
of a more advanced temperature solution. The results are promising and will soon be announced in
a separate paper.

\section{Acknowledgements}

We appreciate the comments about the early version of this paper from
Dr. David Vokrouhlick\'y and Dr. David Nesvorn\'y.
The research was supported by Polish State Committee of Scientific Research grant 1 P03D 020 27.

\bibliographystyle{mn2e}
%\bibliography{paper08}

\appendix

\section[]{Special functions}
\label{apen}

Generally speaking, we use the definitions of special functions according to \citet{BieLo}.
The only exception from this rule is the inclusion of the Condon-Shortley phase $(-1)^m$
in the definition of associated Legendre functions $P_l^m$.
All formulae from other sources have been adjusted to fit this convention.

\subsection{Spherical harmonics}
\label{sferfun}

If $P_l(u)$ stands for the usual Legendre polynamial of degree $l$,
we define the associated Legendre functions (ALF) as
\begin{equation}\label{lf}
    P_l^m(u) = \frac{(-1)^m}{2^l\,l!} \, \left(1-u^2\right)^\frac{m}{2} \,\frac{\mathrm{d}^{l+m}
    (u^2-1)^l}{\mathrm{d} u^{l+m}} = (-1)^m\,\left(1-u^2\right)^\frac{m}{2}\,\frac{\mathrm{d}^m\,P_l(u)}{\mathrm{d}u^m}.
\end{equation}
Introducing the normalization factor $\sigma_{l,m}$
\begin{equation}\label{slm}
  \sigma_{l,m} =  \sqrt{\frac{2l+1}{4 \pi}\,\frac{(l-m)!}{(l+m)!}},
\end{equation}
we define the normalized ALF
\begin{equation}\label{TH}
    \Theta_l^m(u) =  \sigma_{l,m} P_l^m(u).
\end{equation}

Spherical harmonics $Y_{l,m}$ used in this paper are
\begin{equation}\label{sfh}
Y_{l,m}(u,\lambda) =  \Theta_l^m(u)\, e^{i\,m\,\lambda},
\end{equation}
where $u$ is the sine of latitude (i.e., cosine of colatitude) and $\lambda$
is the longitude of the point on a unit sphere.

The symmetries of Legendre functions
\begin{eqnarray}
    P_l^{-m}(u) & = & (-1)^{m} \frac{(l-m)!}{(l+m)!}\,P_l^m(u), \label{plmm}\\
    \Theta_l^{-m}(u) & = & (-1)^{m} \,\Theta_l^m(u), \label{THlmm}
\end{eqnarray}
lead to
\begin{equation}
    Y^\ast_{l,m}(u,\lambda)  =  (-1)^{m} \,Y_{l,-m}(u,\lambda), \label{plmcon}
\end{equation}
where an asterisk indicates a complex conjugate.

The following orthogonality relations hold
\begin{equation}\label{orto}
    \int_{0}^{2\pi} \mathrm{d}\lambda \int_{-1}^1 Y_{j,k}(u,\lambda)\,  Y^\ast_{l,m}(u,\lambda)
   \, \mathrm{d}u  =   (-1)^{m} \int_{0}^{2\pi} \mathrm{d}\lambda \int_{-1}^1 Y_{j,k}(u,\lambda)\,  Y_{l,-m}(u,\lambda)
   \, \mathrm{d}u = \delta_{j,l}\,\delta_{k,m},
\end{equation}
with $\delta_{p,q}$ designating the Kronecker delta function equal to 1 if $p=q$ and zero otherwise.

A function $F(u,\lambda)$ defined on a unit sphere can be expanded in spherical harmonic series
\begin{eqnarray}\label{serh}
    F(u,\lambda)& = & \sum_{l\geqslant 0} \sum_{m = -l}^l a_{l,m}\,Y_{l,m}(u,\lambda), \\
    a_{l,m} & = &
 \int_{-1}^{1} \mathrm{d}u \int_{0}^{2\pi} F(u,\lambda) \,  Y^\ast_{l,m}(u,\lambda)
   \, \mathrm{d}\lambda. \label{acoef}
\end{eqnarray}
If $F$ is real-valued, then the coefficients $a_{l,m}$ admit the property similar to
(\ref{plmcon})
\begin{equation}\label{almcon}
    a_{l,m}^\ast = (-1)^m\,a_{l,-m}.
\end{equation}

\subsection{Wigner functions (generalized spherical harmonics)}
\label{Wigner}

Consider two Cartesian reference frames $Oxyz$ and $Ox'y'z'$ having the same origin $O$,
and related by rotation defined in terms of the 3-2-3 Euler angles $\alpha$, $\beta$, $\gamma$
\begin{equation}\label{rot:ap}
    \vv{r}' = \mtr{R}_3(\gamma) \mtr{R}_2(\beta)
    \mtr{R}_3(\alpha)\, \vv{r},
\end{equation}
with the standard rotation matrices
\begin{equation}\label{R123}
    \mtr{R}_1(\phi) = \left(%
\begin{array}{ccc}
  1 & 0 & 0 \\
  0 &\cos{\phi} &  \sin{\phi}   \\
  0 &- \sin{\phi} & \cos{\phi}   \\
\end{array}%
\right), \quad
    \mtr{R}_2(\phi) = \left(%
\begin{array}{ccc}
  \cos{\phi} & 0 & -\sin{\phi}   \\
  0 & 1 & 0 \\
  \sin{\phi} &0& \cos{\phi}   \\
\end{array}%
\right), \quad
    \mtr{R}_3(\phi) = \left(%
\begin{array}{ccc}
  \cos{\phi} &  \sin{\phi} & 0 \\
  - \sin{\phi} & \cos{\phi} & 0 \\
  0 & 0 & 1 \\
\end{array}%
\right).
\end{equation}
The same transformation can also be specified in terms of more common 3-1-3 Euler angles
as\footnote{\textsf{corrected}}
\begin{equation}\label{rot:313}
    \vv{r}' = \mtr{R}_3(\gamma- \pi/2) \mtr{R}_1(\beta) \mtr{R}_3(\alpha+\pi/2)\, \vv{r}.
\end{equation}

According to the transformation laws of spherical harmonics for the 3-2-3 Euler angles
sequence, each spherical function of polar coordinates $(\theta, \lambda)$ in $Oxyz$ frame
becomes a combination of the same degree spherical functions
of polar coordinates $\theta'$ and $\lambda'$ in $Ox'y'z'$. Thus, for each harmonic
\begin{equation}\label{tran:inv}
  Y_{l,m}(\cos{\theta'},\,\lambda')   = \sum_{k=-l}^l
     D^l_{k,m}(\alpha,\beta,\gamma) \,Y_{l,k}(\cos{\theta},\,\lambda),
\end{equation}
or, conversely,
\begin{equation}\label{tran1}
    Y_{l,m}(\cos{\theta},\,\lambda) = \sum_{k=-l}^l
   \left[ D^l_{m,k}(\alpha,\beta,\gamma)\right]^\ast\,Y_{l,k}(\cos{\theta'},\,\lambda'),
\end{equation}
where the general definition of Wigner D-matrix elements (or Wigner D-functions)
is \citep{BieLo}
\begin{equation}\label{Wig1}
    D^l_{m,k}(\alpha,\beta,\gamma)  =  d^l_{m,k}(\beta) \,e^{-i(m \alpha+k \gamma)}.
\end{equation}
The symbols $d^l_{m,k}(\beta)$ designate so called Wigner d-functions
\begin{equation}
    d^l_{m,k}(\beta)  =  \sqrt{\frac{(l+k)! (l-k)!}{(l+m)! (l-m)!}}\,  \left[\sin{(\beta/2)}\right]^{k-m} \left[\cos{(\beta/2)}\right]^{k+m}\,
    P_{l-k}^{(k-m,k+m)}(\cos{\beta}). \label{smald}
\end{equation}
The definition of d-functions in Eq.~(\ref{smald}) involves the Jacobi polynomials
\begin{equation}
    P_n^{(a,b)}(x)  =
    \sum_{m=0}^n \left(%
\begin{array}{c}
  n+a \\
  m \\
\end{array}%
\right) \left(%
\begin{array}{c}
  n+b \\
  n-m \\
\end{array}%
\right)\,\left(
\frac{x-1}{2}\right)^{n-m}\,\left(
\frac{x+1}{2}\right)^{m}. \label{Jac}
\end{equation}
Wigner d-functions $d^l_{m,k}$ admit some parity properties like
\begin{equation}\label{par}
    d^l_{m,k}   =  (-1)^{k+m} d^l_{-m,-k}, \qquad
    d^l_{m,k}   =  (-1)^{k+m} d^l_{k,m}.
\end{equation}
Using these two identities, one can always reduce the $d^l_{m,k}$ function to the
case when $k+m \geq 0$ and $k-m \geq 0$, avoiding the appearance of negative powers
of sines and cosines in Eq.~(\ref{smald}).

As it follows from (\ref{par}), the conjugate of $D^{l}_{m,k}$ is
\begin{equation}\label{Dcon}
    \left[D^{l}_{m,k}(\alpha,\beta,\gamma)\right]^\ast = (-1)^{m-k} D^{l}_{-m,-k}(\alpha,\beta,\gamma).
\end{equation}

In the special case $d^l_{k,0}$ or $d^l_{0,k}$, Wigner d-functions reduce to AFL
\begin{equation}\label{dl0k}
  d^l_{k,0}(\beta) = (-1)^k\, d^l_{0,k}(\beta) = \sqrt{\frac{(l-k)!}{(l+k)!}}\,P_l^k(\cos{\beta})
  = \sqrt{\frac{4\,\pi}{2l+1}}\,\Theta_l^k(\cos{\beta}).
\end{equation}

\subsection{Overlap integrals and Gaunt coefficients}
\label{OVG}

The Wigner 3-j symbols are numerical coefficients defined as a sum
\begin{eqnarray}\label{3j}
 \left[%
\begin{array}{ccc}
  j_1 & j_2 & j_3 \\
  m_1 & m_2 & m_3 \\
\end{array}%
\right] & = & \left[ \frac{(j_1+j_2-j_3)!\,(j_1-j_2+j_3)!\,(-j_1+j_2+j_3)!}{(j_1+j_2+j_3+1)!}\right]^\frac{1}{2}
\sqrt{(j_1+m_1)!\,(j_1-m_1)! \,(j_2+m_2)!\,(j_2-m_2)! \,(j_3+m_3)!\,(j_3-m_3)! } \nonumber \\
& & \sum_k \frac{(-1)^{k+j_1-j_2-m_3}}{k!\,(j_1+j_2-j_3-k)!\,(j_1-m_1-k)!\,(j_2+m_2-k)!\,(j_3-j_2+m_1+k)!\,(j_3-j_1-m_2+k)!},
\end{eqnarray}
where the index $k$ takes all integer values leading to nonnegative factorial arguments.

The so-called overlap integral is an integral of the product of two associated Legendre
functions with the integration limits $\pm1$. We use a normalized version of this function
\begin{equation} \label{over}
   w_{n_1,n_2}^{m_1,m_2} = \int_{-1}^1 \Theta_{n_1}^{m_1}(u) \, \Theta_{n_2}^{m_2}(u)\,\mathrm{d}u.
\end{equation}
Using a formula of \citet{Dong:02}, converted to the normalized
version, we can express the overlap integrals in terms of the Wigner 3-j symbols
\begin{eqnarray}
  w_{n_1,n_2}^{m_1,m_2} &=& \frac{(-1)^{\min(m1,m2)}}{\pi} \, 2^{|m_1-m_2|-4}\,|m_1-m_2|
  \sqrt{(2n_1+1)(2n_2+1)} \nonumber \\
   & & \sum_{l=\max(|n_1-n_2|,|m_1-m_2|)}^{n_1+n_2} (2l+1)\, G(|m_1-m_2|,l) \,  \left[%
\begin{array}{ccc}
  n_1 & n_2 & l \\
  0 & 0 & 0 \\
\end{array}%
\right]    \left[%
\begin{array}{ccc}
  n_1 & n_2 & l \\
  -m_1 & m_2 & m_1-m_2 \\
\end{array}%
\right] \\
G(m,l) & = & \left(1+ (-1)^{l+m}\right) \sqrt{\frac{(l-m)!}{(l+m)!}} \frac{\Gamma(l/2)\,\Gamma((l+m+1)/2)}{((l-m)/2)!\,\Gamma((l+3)/2)}.
\end{eqnarray}
Obviously, the overlap integrals are symmetric with respect to the indices $w_{n_1,n_2}^{m_1,m_2}=
w_{n_2,n_1}^{m_2,m_1}$, and they vanish either if $|m_i| > n_i$, or if the integrand in Eq.~(\ref{over})
is an odd function of $u$, i.e., when $(-1)^{n_1+n_2+m_1+m_2} = -1$. Moreover, because of the orthogonality of
the associated Legendre functions, we have $w_{n_1,n_2}^{m,m}= 0$, for $n_1 \neq n_2$.

Because of a frequent appearance of $w_{1,2p}^{1,0}\,\sigma_{1,1}^{-1}$ in the final formulae
we introduce the special case of the overlap integral
\begin{equation}\label{W2p}
    W_{p} = \frac{w_{1,2p}^{1,0}}{\sigma_{1,1}} = \frac{\sqrt{\pi}}{4^{p+1}}
    \left( \frac{(2p-1)!!}{p!} \right)^2 \frac{\sqrt{4p+1}}{(p+1)(2p-1)} = \frac{\sqrt{\pi\,(4p+1)}}{4\,(p+1)(2p-1)}\,\left[P_{2p}(0)\right]^2.
\end{equation}

The Gaunt coefficients (or Gaunt integrals) are defined as an integral of three spherical functions over the unit sphere
surface
\begin{equation}\label{G}
    \mathcal{G}_{l_1,l_2,l_3}^{m_1,m_2,m_3} = \int_{-1}^1 \mathrm{d}u \int_{0}^{2 \pi}
    Y_{l_1,m_1}(u,\lambda)\,Y_{l_2,m_2}(u,\lambda)\,Y_{l_3,m_3}(u,\lambda)\,\mathrm{d}\lambda.
\end{equation}
Their expression in terms of the Wigner 3-j symbols is much simpler than for the overlap integrals
\begin{equation}\label{G3j}
 \mathcal{G}_{l_1,l_2,l_3}^{m_1,m_2,m_3} = \sqrt{\frac{ (2l_1+1)(2l_2+1)(2l_3+1)}{4\pi}}
 \left[%
\begin{array}{ccc}
  l_1 & l_2 & l_3 \\
  0 & 0 & 0 \\
\end{array}%
\right] \,  \left[%
\begin{array}{ccc}
  l_1 & l_2 & l_3 \\
  m_1 & m_2 & m_3 \\
\end{array}%
\right].
\end{equation}
Gaunt coefficients are non-zero if the following three conditions
are satisfied:
\begin{equation} \label{G:prop}
  (-1)^{l_1+l_2+l_3} = 1, \quad
  m_1+m_2+m_3 = 0, \quad  |l_i-l_j| \leqslant l_k  \leqslant l_i+l_j.
\end{equation}
Note, that the signs of all $m_i$ superscripts can be simultaneously inverted, and the pairs $l_i,m_i$ can be arbitrarily permuted
\begin{equation}\label{Ginv}
 \mathcal{G}_{l_1,l_2,l_3}^{m_1,m_2,m_3} = \mathcal{G}_{l_1,l_2,l_3}^{-m_1,-m_2,-m_3} = \mathcal{G}_{l_2,l_1,l_3}^{m_2,m_1,m_3}
 = \mathcal{G}_{l_3,l_1,l_2}^{m_3,m_1,m_2}, ~\mbox{etc.}
\end{equation}
From the three-term recurrence identities related with the Wigner 3-j symbols \citep{LuLu:98}, we can deduce useful relation
\begin{equation}\label{Grec}
 \left( l_3 (l_3+1)-l_2\,(l_2+1)\,-l_1\,(l_1+1) \right)\,\mathcal{G}_{l_1,l_2,l_3}^{m,0,-m} =
 \sqrt{l_2\,(l_2+1)}\,\left[\sqrt{(l_1-m)(l_1+m+1)} \mathcal{G}_{l_1,l_2,l_3}^{m+1,-1,-m} +
 \sqrt{(l_1+m)(l_1-m+1)} \mathcal{G}_{l_1,l_2,l_3}^{m-1,+1,-m}\right].
\end{equation}
The Gaunt coefficients appear in the transformation rule for the product of spherical harmonics (Clebsch-Gordan series).
Applying Eq.~(\ref{G}) to (\ref{acoef}) we find
\begin{equation}\label{Y:prod}
 Y_{j_1,k_1} Y_{j_2,k_2} = \sum_{l = \max{(|j_1-j_2|,|k_1+k_2|)}}^{j_1+j_2} (-1)^{k_1+k_2} \mathcal{G}_{j_1,j_2,l}^{k_1,k_2,-k1-k2} \,Y_{l,k_1+k_2},
\end{equation}
where the sum $j_1+j_2+l$ must be even.

\label{lastpage}
\end{document}